\documentclass[%
 amsmath,amssymb,
 aps,
 prfluids,
]{revtex4-2}

\usepackage{graphicx}
\usepackage{dcolumn}
\usepackage{bm}

\usepackage{lipsum}
\usepackage{ulem}
\usepackage{enumerate}

 \usepackage[table,xcdraw]{xcolor}

 \newcommand{\review}[1]{\textcolor{black}{#1}}
 \newcommand{\reviewTwo}[1]{\textcolor{black}{#1}}

\newcommand{\eref}[2][]{(\ref{#2}#1)}
\renewcommand{\emph}[1]{\textit{#1}}
\begin{document}

\preprint{APS/123-QED}

\title{{Subcritical axisymmetric solutions in rotor-stator flow}}


\author{Artur Gesla}
 \altaffiliation[Also at ]{LISN-CNRS, Universit\'e Paris-Saclay, F-91400 Orsay, France}
\affiliation{%
Sorbonne Universit\'e, F-75005 Paris, France}%




\author{Yohann Duguet}
\author{Patrick Le Qu\'er\'e}
\affiliation{%
 LISN-CNRS, Universit\'e Paris-Saclay, F-91400 Orsay, France
}%
\author{Laurent Martin Witkowski}%
\affiliation{%
 Universite Claude Bernard Lyon 1, CNRS, Ecole Centrale de Lyon, INSA Lyon, LMFA, UMR5509, 69622 Villeurbanne, France
}%


\date{\today}


\begin{abstract} 
Rotor-stator cavity flows are known to exhibit unsteady flow structures in the form of circular and spiral rolls. While the origin of the spirals is well understood, that of the circular rolls is not. In the present study the axisymmetric flow in an aspect ratio $R/H=10$ cavity is revisited {numerically using recent concepts and tools from bifurcation theory}. It is confirmed that a linear instability takes place at a finite critical Reynolds number $Re=Re_c$, and that there exists a subcritical branch of large amplitude chaotic solutions. This motivates the search for subcritical finite-amplitude solutions. The branch of periodic states born in a Hopf bifurcation at $Re=Re_c$, identified using a Self-Consistent Method (SCM) and arclength continuation, is found to be supercritical. The associated solutions only exist, however, in a very narrow range of $Re$ and do not explain the subcritical chaotic rolls.
Another subcritical branch of periodic solutions is found using the Harmonic Balance Method with an initial guess obtained by SCM. In addition, edge states separating the steady laminar and chaotic regimes are identified using a bisection algorithm. These edge states are bi-periodic in time for most values of $Re$, {where} their dynamics is {analysed in detail}. Both solution branches fold around at approximately the same value of $Re$, which is lower than $Re_c$ yet still larger than the values reported in experiments. This suggests that, at least in the absence of external forcing, sustained chaotic rolls  have their origin in the bifurcations from these unstable solutions.

\end{abstract}

\maketitle


\section{Introduction}
Rotating flows in closed containers
have motivated many hydrodynamic stability studies 
due their simplicity, their relevance to various industrial configurations and the dramatic patterns
observed \cite{Moisy_epn_03,Launder_arfm_10}. Hydrodynamic stability theory itself has long rested on the difference between 
open and closed flow geometries, which leads respectively to the opposition
between local and global methods of analysis \cite{Dauchot_jp2_1997}. 
The case of \emph{\large} cylindrical containers, with a finite radius much larger than their height,
falls somewhat in between, and it is not always clear which kind of analysis is most relevant: global mode analysis, where all modes are consistent with the boundary conditions, or the (often simpler) local analysis of perturbations as in infinitely extended systems. We consider in this study the incompressible flow between a rotor and a stator, in the case where the lateral circular wall rotates together with the rotor. This flow has been investigated in the late 1990s in detail, mainly using experimental facilities \cite{Lopez_pof_1996,Schouveiler_jfm_2001,Gauthier_jfm_1999,Poncet_pof_2009} but also later two-dimensional and three-dimensional numerical computations \cite{Daube_cf_2002,Poncet_pof_2009,Lopez_pof_2009,Serre_pof_04,Lopez_pof_1996,Gelfgat_fdr_2015}. The flow is characterised by two independent parameters~: a Reynolds number $Re$ proportional to the rotor’s angular velocity $\Omega$, and a geometrical aspect ratio $\Gamma=R/H$ where $H$ and $R$ are the respective height and radius of the cylinder. The base flow for the configuration of interest here, where $\Gamma \ge 5$, is axisymmetric.  It has a dominantly azimuthal flow with strong shear at the walls, as well as a meridional recirculation, associated, for high enough $Re$, with two separated boundary layers, one along the stator (labelled as the Bödewadt boundary layer) and the other along the rotor (labelled as the Ekman boundary layer).  \\


The linear stability of this axisymmetric flow with respect to three-dimensional perturbations was carried out by Gelfgat \cite{Gelfgat_fdr_2015}. The predictions are in very good agreement with experiments \cite{Schouveiler_pof_1998,Gauthier_jfm_1999} as well as with direct numerical simulations \cite{Peres_jcp_2012}.
The linear instability from which spiral patterns originate \cite{Gelfgat_fdr_2015} is a supercritical Hopf bifurcation. The resulting patterns saturate in amplitude and rotate around the cavity at an angular velocity smaller than that of the rotor. Yet, although  well understood these beautiful spiral patterns are not the earliest manifestation of unsteadiness encountered as $Re$ is increased from zero. As reported in most experimental studies so far, \emph{axisymmetric ring-like perturbations} are also observed. Viewed in a meridional plane they correspond to wavepackets of vortices located near the stator.
At low enough $Re$ close to their experimental onset they are simply convected inwards and disappear as they pass near the axis before they reach the rotor \cite{Gauthier_jfm_1999}. At higher $Re$
the rolls described in \cite{Schouveiler_jfm_2001} form a steady front with complex internal dynamics. Adopting a local-like description, these rings are usually described as emerging from a convective instability of the Bödewadt layer, developing from the corner, and propagating inwards along the stator until the axis region where they are damped \cite{Poncet_pof_2009}. Note that for larger $Re$, there is a mix of spiral and circular rolls. As far as experiments and numerics can be quantitatively compared, an axisymmetric instability was reported in Ref.~\cite{Daube_cf_2002}
 at $Re \approx 2950$. This has to be contrasted with the experimental values: rings are reported for $Re$ as low as 180 \cite{Gauthier_jfm_1999} and 160 \cite{Schouveiler_jfm_2001} for a similar aspect ratio
(using the same definition of $Re$). The configuration with $\Gamma=5$ has been also simulated using accurate spectral codes \cite{Lopez_pof_2009,Serre_jfm_2001}, \review{although the circular patterns reported in these studies are transients}. 
The simulated ring patterns are more clearly sustained in case the rotation of the disc is temporally modulated \cite{do2010optimal}. It is hence tempting to interpret them as a (linear) response to external forcing, in other words \emph{not} as a genuine nonlinear state of the system. Refs~\cite{Lopez_pof_2009,do2010optimal} as well as \cite{Poncet_pof_2009} support the view that these vortical structures are only transient responses to external effects. The overlooked investigation by Daube and Le Qu\'er\'e \cite{Daube_cf_2002} goes beyond by identifying, now for $\Gamma=10$, three new properties~: i)~a~linear instability respecting the axisymmetry of the flow was identified using eigenvalue, at a value of $Re$ above that where rings can be observed, ii) quantitatively huge levels of transient growth before iii) the ring patterns saturate in amplitude, {iv)~the existence of a subcritical large amplitude chaotic solution branch}. These new features, especially i) and ii) are familiar to the subcritical shear flow community working on plane Poiseuille and boundary layer flows \cite{manneville2016transition,eckhardt2018transition}.\\

Several questions can be asked at this point : is the critical point of Daube and Le Qu\'er\'e confirmed and how robust is it? Does a branch of {axisymmetric nonlinear states} bifurcate subcritically from this point? Is it physically consistent with the convective instability put forward by experimentalists? 
{The lack of radial propagation of the associated envelope, attested from experimental space-time diagram, suggests that nonlinearity can enter the picture. Adopting such a point of view, the quasi-steady envelopes of rings} could be interpreted as genuinely nonlinear states, with a complex time dependency but a coherent spatial structure. This should be contrasted with the concept of noise-sustained coherent structure \cite{tobias1998convective,Poncet_pof_2009}. The subcriticality of the ring regime and the presence of an axisymmetric instability threshold, both reported in Ref. \cite{Daube_cf_2002}, would make the nonlinear branch emerging from the critical point, if it is confirmed to also bifurcate subcritically, the ideal candidate to explain experimental observations. Subcriticality {would also imply} that\review{ , for $Re<Re_c$, } only finite-amplitude perturbations above a certain threshold can excite this non-trivial regime.\\


In order to shed a light on the interrogations above, the present numerical work aims at better characterising the stability threshold and its numerical robustness, at identifying for the first time nonlinear branches of solutions possibly connected to the critical point, and at debating whether the corresponding nonlinear states bear any relation with the patterns observed experimentally {and numerically}. Among all possible nonlinear states that can exist beyond the laminar base flow, many nonlinear solutions of the subcritical regime are unstable. Some of them  possess the property of lying on the laminar-turbulent boundary, the so-called \emph{edge} manifold in the associated state space. Numerical simulations constrained to lie in that manifold converge asymptotically to specific equilibrium regimes called \emph{edge states}. Their determination is specifically interesting for subcritical transition since their instability captures the mechanisms leading from finite-amplitude initial perturbations of the base flow towards the turbulent state. We focus here on the same \review{flow, restricted to the axisymmetric subspace, in a cavity of} aspect ratio $\Gamma=10$ as in Ref. \cite{Daube_cf_2002}. As will become clear throughout the paper, many traditional numerical algorithms {turn out to be} technically too limited for the search of nonlinear states. This represents an opportunity to improve them or to use recent methods whose relevance will be {also} discussed.\\


The paper is structured according to the following plan : section \ref{num-mtds} describes briefly the numerical methods used, section \ref{sec-base} describes the laminar base flow and its stability, \ref{sec-per-sol} describes the nonlinear periodic states found in the flow, section \ref{sec-bis-branch} describes the edge tracking and the edge state {dynamics} and finally section \ref{sec-summary} summarises the results and gives outlooks for future work.

\section{Numerical methods} \label{num-mtds}
\subsection{Flow configuration}

\begin{figure}
    \centering
    \includegraphics[width=0.5\columnwidth]{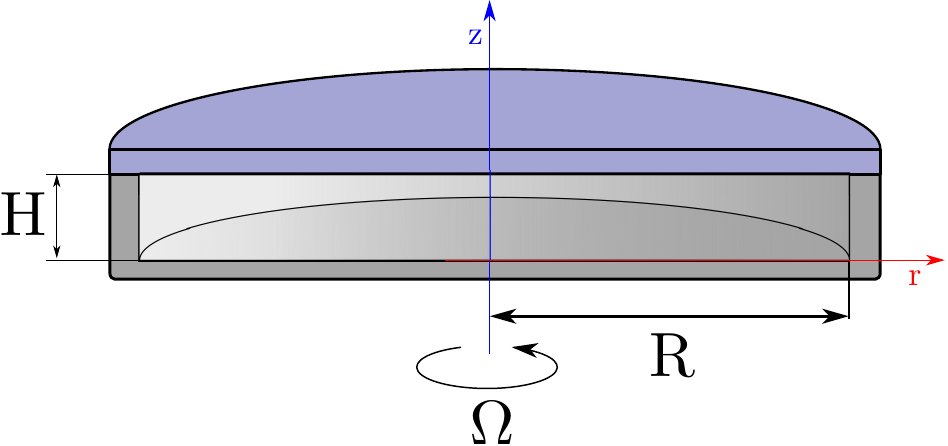}
    \caption{Rotor-stator geometry with rotating shroud. The set-up is characterised by the Reynolds number $Re= {\Omega H^2}/{\nu}$ and an aspect ratio $\Gamma=R/H$ fixed here to 10.}
    \label{rs}
\end{figure}

The system under consideration is an axisymmetric rotor-stator flow. The two-dimensional meridional geometry is depicted in
Figure \ref{rs}. The flow is governed by the incompressible Navier-Stokes equations \eref[a]{ns} and \eref[b]{ns} {written in polar coordinates{ $(r,z)$}, with the velocity  $\mathbf{u}$ characterized by its three components ($u_r,u_\theta$, $u_z$)} :
\begin{subequations} \label{ns}
\begin{align}
\frac{\partial \mathbf{u}}{\partial t}+( \mathbf{u}\cdot \nabla) \mathbf{u}&=-\frac{1}{\rho}  \mathbf{\nabla} p + \nu \mathbf{\nabla}^2  \mathbf{u} \\
 \mathbf{\nabla} \cdot  \mathbf{u}&=0
\end{align}
\end{subequations}
together with the boundary conditions \eref{bc}
\review{
\begin{equation} \label{bc}
\begin{cases}
    {\bm u}=r\Omega{\bm e_{\theta}},& \text{at the rotor $(z=0)$} \\
    {\bm u}=R\Omega {\bm e_{\theta}},& \text{at the shroud $(r=R)$} \\
    {\bm u}={\bm 0},& \text{on the stator $(z=H)$}, \\
  \end{cases}
\end{equation}
}
 {where $\Omega$ is the angular frequency of the disc rotation, $\nu$ is the kinematic viscosity of the fluid and $\rho$ its density. As in \cite{Daube_cf_2002}, the geometry is characterised by a radius-to-gap ratio $\Gamma=R/H$ equal to 10. 
 {Non-dimensionalisation of all lengths, and of the velocity field $\mathbf{u}$ is achieved respectively using the lengthscale $H$ and velocity scale $\Omega H$, without change of notation.}} Different possible definitions for the Reynolds number $Re$ include $Re_R=\Omega R^2/ \nu$, built on the length scale $R$, $\Omega H^{3/2}R^{1/2}/ \nu$ when the centrifugal force is the key destabilizing mechanism as it is the case in exactly counter-rotating disks \cite{Witkowski_jfm_2006}, or $Re_{RH}={\Omega R H}/{\nu}$. 
In this study $Re_H=\Omega H^2/\nu$, related to the length scale $H$, is chosen.
\subsection{Spatial discretisation}
The continuous problem is discretised using a second order Finite Volume method on a staggered grid. The details of the discretisation as well as the staggered arrangement can be found, for instance, in {appendix A of} Ref. \cite{Faugaret_jfm_2022}. Two types of meshes, uniform and non-uniform, are used in the present study. The uniform mesh on one hand allows for direct, reproducible and easy comparison with the existing literature. The non-uniform mesh, on the other hand, allows for greater accuracy 
at the price of only a moderate increase in computational cost. The non-uniform mesh is refined in the regions near the rotor, the stator and  the rotating shroud. In the radial direction it is uniform from $r=0$ to $r=8$ and then refined on the remaining $r\in(8,10)$ according to the formula 
\begin{subequations} \label{meshr}
\begin{align}
    b_i=\frac{1}{2}\left(1+\frac{tanh(\delta(x_i-\frac{1}{2}))}{tanh(\delta/2)}\right) \\
    r_i=8+2\cdot\frac{b_i}{(a+(1-a) b_i)}    
    \end{align}
\end{subequations}
{with $\delta= 0.7258$ and $a=0.4989$, where $x_i\in(0,1)$ is the uniform mesh.} 70\% of total number of $N_r$ grid points are used in the uniform region $r\in(0,8)$ and the remaining 30\% are used in the non-uniform region $r\in(8,10)$.  The non-uniform mesh in the axial direction follows the formula
\begin{subequations} \label{meshz}
\begin{align}
    z_i=\frac{1}{2}\left(1+\frac{tanh(\delta(\frac{i}{N_z}-\frac{1}{2}))}{tanh(\delta/2)}\right) 
    \end{align}
\end{subequations}
   with $ \delta=2.8587 $. Details on the resulting cell size for the pressure are given for two mesh types in Table \ref{tab-mesh}. In both the uniform and non-uniform case, the number of points in the radial direction $r$ (resp. the axial direction $z$) is noted $N_r$ (resp. $N_z$).

\begin{table}[]
    \centering
    \begin{tabular}{c|c|cccc}
        {$\ N_r \times N_z$} & type & max $dr$ & min $dr$ & max $dz$ & min $dz$ \\ \hline
         \ 600 $\times$ 160 & uniform & 0.0167 & --- & 0.0062 & --- \\
         1024 $\times$ 192 & non-uniform & 0.0114 & 0.0028 & 0.0083 & 0.0017 
    \end{tabular}
    \caption{Uniform (600$\times$160 cells in $r$-$z$) and non-uniform (1024$\times$192 cells in $r$-$z$) mesh used in the study. $dr$ and $dz$ correspond to the horizontal and vertical dimension of the pressure cell in the finite volume discretisation. 
    }
    \label{tab-mesh}
\end{table}

\subsection{Numerical simulation} \label{num-sim}
 Time integration is carried out using a Backwards Differentiation Formula 2 scheme (BDF2). It uses a prediction-projection algorithm in the rotational pressure correction formulation, as described in \cite{Guermond_cmame_2006}. {The diffusion term treated implicitly gives rise to a Helmholtz problem for each velocity component increment. These Helmholtz problems are solved using an alternating-direction implicit (ADI) method in incremental form which preserves the second order accuracy in time. In the projection step the velocity field is projected onto the space of divergence free fields by solving a Poisson equation for pressure. This Poisson equation is solved using a direct sparse solver. Time integration is performed with a time step $dt$ corresponding to a Courant–Friedrichs–Lewy (CFL) number of 0.3. }
 

The nonlinear system of Eqs. \eref{ns} and \eref{bc} admits for all $Re$ a steady  solution also called $({\bm U}_b,P_b)$. Once the system is discretised, the solution of the large algebraic nonlinear system of equations is determined numerically using a Newton-Raphson algorithm. The $O(4N_rN_z)$ unknowns are the velocity and pressure values at each discretisation point. The linear stability of the base flow is evaluated using the Arnoldi method based on a well-validated ARPACK package \cite{Lehoucq_1998}.
\review{
The shift-and-invert mode is used to find the interesting subset of the eigenvalues. 
For a generalised eigenvalue problem
\begin{equation}
    \bm{A} \bm{w}=\lambda \bm{B} \bm{w},
\end{equation}
the shift-and-invert method finds a subset of eigenvalues closest to a complex shift $s$ by repeated Arnoldi iteration:\begin{equation}
    \nu \bm{w}=\left( \bm{A}-s \bm{B} \right)^{-1}\bm{B}\bm{w}
\end{equation}
where the original eigenvalues $\lambda$ can be retrieved  with:
\begin{equation}
    \nu=\frac{1}{\lambda-s}
\end{equation}
}
We take advantage of the Finite Volume discretisation which leads to sparse matrices. Building and storing explicitly the matrix associated with the Jacobian matrix is not a problem since the number of non-zero elements {scales with} the number of grid points. 
Most techniques used to compute the base flow and the associated eigenmodes are similar to those presented in appendix B of \cite{Faugaret_jfm_2022}. The main change is the use of MUMPS~\cite{Amestoy_siamjmaa_2001} as the direct solver for the linear system at each iteration step of Newton-Raphson algorithm. Due to the presence of the thin boundary layers close to each disk, the set of equations resulting from the discretisation of the continuous system is in general poorly conditioned. Sparse direct solvers will be therefore preferred over iterative solvers. Once $({\bm U}_b,P_b)$ is found, the time-stepping code can be adapted to evolve the perturbation to the base flow rather than the full velocity field itself, at the expense of a triple evaluation of the convective terms. 

Two recently developed methods are additionally used in this study : the Self-Consistent Method (SCM) introduced in \cite{mantic_prl_2014} and the Newton method for converging periodic orbits, similar to the algorithm described in \cite{bengana_prf_2021} and \cite{SierraAusin_cmame_2022}. Those will be briefly explained in section \ref{sec-per-sol}.


\section{Stability of the base flow \label{sec-base}}

\subsection{Base flow and critical Reynolds number}
The first step in the analysis of the present flow case is to identify numerically the steady base flow and to study its stability as a function of the Reynolds number $Re$. 
The base flow $(\mathbf{U}_b,P_b)$ is shown for $Re=3000$ in Figure \ref{evc}. The base flow features, as expected, shear due to axial gradients of azimuthal velocity, and a meridional recirculation. The 
\review{structure}
of the base flow has been investigated in detail in Ref. \cite{Cousin_cras_99}.
{For high enough $Re$, the base flow is formally self-similar in an $O(1)$ region around the axis, with the bulk in solid body rotation, with a rotation velocity $0.313$.
A snapshot of the azimuthal velocity is shown (in perturbation mode) for the chaotic rolls at the same value of $Re$ in Figure \ref{evc}c). The vortical structures are mainly located between $r=1$ and $6$. Occasional ejections of fluid {from the B\"odewadt layer can be seen, for instance in Figure \ref{evc}c) for $4<r<6$}.
\begin{figure}[h]
\begin{tabular}{ll}
a) & \\
 & \includegraphics[width=0.97\columnwidth]{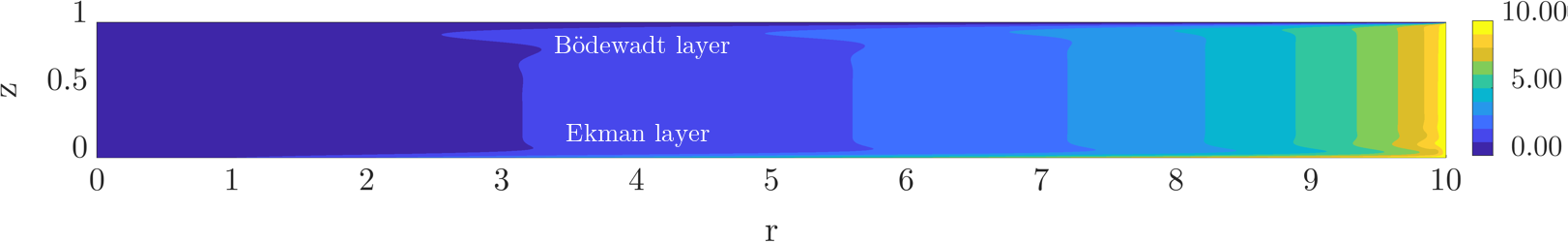}   \\
b) & \\
 & ~\hspace{0.13cm} \includegraphics[width=0.89\columnwidth]{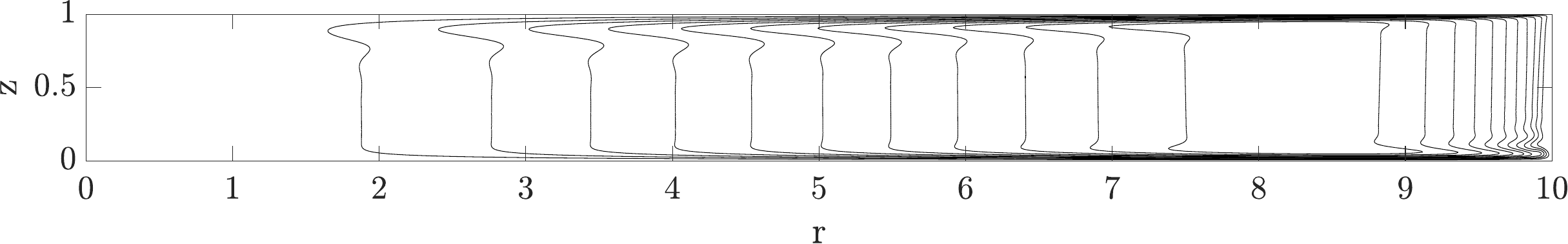} \\
c) & \\
& \includegraphics[width=0.97\columnwidth]{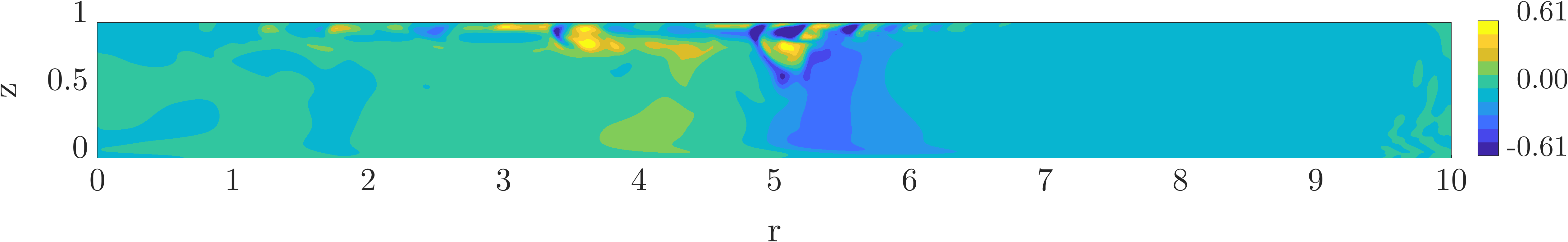} \\
\end{tabular}
    \caption{{a) Azimuthal velocity for the base flow, b) streamfunction associated with the meridional base flow for $\psi = 0$ at the wall to $0.6$ with increments for isovalues of $0.05$, c) instantaneous azimuthal velocity perturbation for chaotic rolls.} $Re=3000$.
    The rotating walls are on the right and at the bottom, the stationary wall at the top, while the left boundary corresponds to the symmetry axis.
The streamfunction $\psi(r,z)$ is defined implicitly by $u_r=\frac{1}{r}\frac{\partial\psi}{\partial z}$ and $u_z=-\frac{1}{r}\frac{\partial\psi}{\partial r}$.}
    \label{evc}
\end{figure}


 The stability of the base flow is studied by introducing infinitesimal perturbation to the velocity and pressure fields in the following form:
      \begin{equation}
    \mathbf{u}=\mathbf{U}_b+\mathbf{u}'e^{\lambda t } +\mathbf{u}'^*e^{\lambda^* t }\ \ \ p=P_b+p'e^{\lambda t } + p'^*e^{\lambda^* t }
      \end{equation}
       {The asterisk denotes complex conjugate.} Plugging this \emph{ansatz} into the Navier-Stokes equations and linearising around  ($\mathbf{U}_b,P_b)$ yields the new system
       \begin{subequations} \label{evp}
           \begin{align}
            \lambda \mathbf{u}' +(\mathbf{u}'\cdot \nabla)\mathbf{U}_b+(\mathbf{U}_b\cdot \nabla)\mathbf{u}' & = - \nabla p' + \frac{1}{Re}\nabla^2 \mathbf{u}' \\
            \nabla \cdot \mathbf{u}' &= 0
           \end{align}
       \end{subequations}
After spatial discretisation the above system can be re-cast into the generic form
 \review{      \begin{equation} \label{ev}
        \textbf{J}\mathbf{q}'=\lambda \mathbf{B}\mathbf{q}'
    \end{equation}
where $\mathbf{q'}=(u_r',u_{\theta}',u_z',p')$,} $\mathbf{J}$ is the Jacobian evaluated on the base flow $(\mathbf{U}_b,P_b)$ and $\mathbf{B}$ is the identity matrix except for the rows corresponding to the continuity equation. The eigenvalues $\lambda \ (= \lambda_r + i \, \lambda_i) $ of the generalised eigenproblem \eref{ev} are found numerically using the ARPACK library in the shift-invert mode. For each shift $\sigma=0.1 \, i,\ 0.5 \, i,\ 1.0 \, i,\ 2.0 \, i$, a fixed number of 200 of eigenvalues is found. {The mesh used in the above example was chosen {uniform with $N_r \times N_z$ =} 600$\times$160 as in \cite{Daube_cf_2002}.} Part of the resulting spectrum is shown in Figure \ref{spec} {for two values of $Re=2900$ and $3000$, respectively below and above the instability threshold.} The base flow and the associated least stable eigenvector for $Re=3000$ are visualised in Figures \ref{evc}a and \ref{evc2}(a-b), respectively. This eigenvector consists {mainly of a steady wavepacket of} counter-rotating rolls localised inside the Bödewadt layer. {The largest amplitude of these rolls corresponds to $1 \le r \le 2$.}
{Weaker yet larger-scale wave-like structures can also be observed outside the Bödewadt layer, displaying various inclinations depending on the radial position. They are interpreted as eigen-oscillations due to the rotation of the core, like inertial waves (whose wavevector's orientation is directly dependent on the ratio between angular frequency and rotation rate) except that the rotation rate varies spatially, like in other non-confined vortical flows \cite{fabre2006kelvin}. Both the rolls and the wave-like structure make up the oscillatory eigenmode. For clearer evidence about the relevance of inertial oscillations to the eigenmodes, we illustrate in Figure \ref{evc2}c another eigenmode from the spectrum at $Re=3000$, with angular frequency $\lambda_i \approx 0.5$ smaller than for the leading eigenmode. This eigenmode does {not} feature any roll structure inside the B\"odewadt layer but features energetic oblique structures with a tighter angle consistently with the expected behaviour for inertial waves. A more detailed investigation of inertial wave contributions to enclosed rotating flows is left for a specific study. }


\begin{figure}
    \centering
    \vspace{2mm}
    \includegraphics[width=0.5\columnwidth]{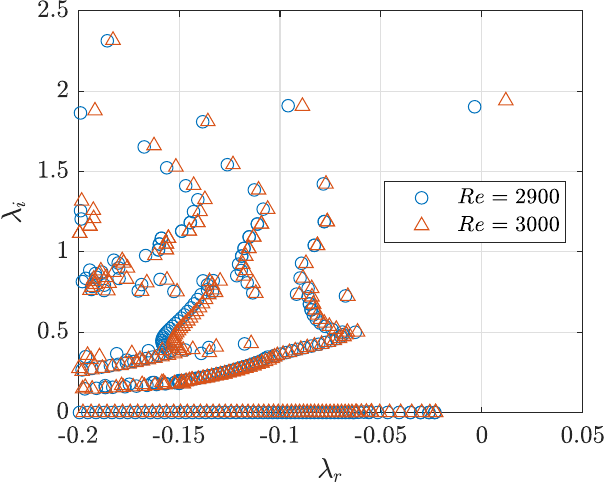}
    \caption{Spectrum of the linearised Navier-Stokes operator in the generalised eigenvalue problem \eref{ev} for two values of $Re=2900$ and $Re=3000$ straddling $Re_c$. The eigenvalue crossing the imaginary axis {defines} the unstable eigenmode of the flow. Spatial resolution R0 (see Table \ref{reolutions}).}
    \label{spec}
\end{figure}

\begin{figure}
    \centering
        
          \includegraphics[width=\columnwidth]{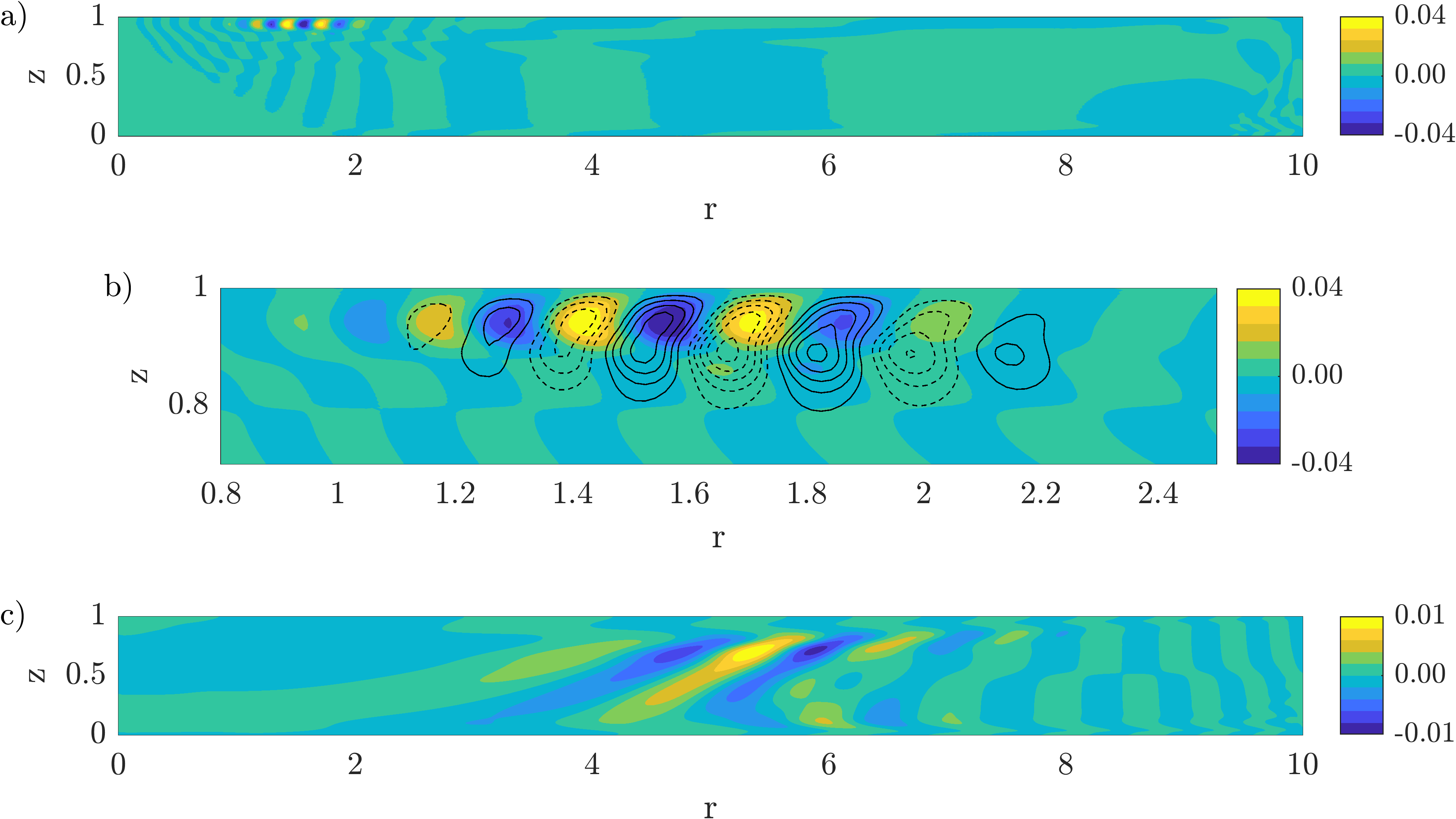}

    \caption{a) Azimuthal velocity component of the unstable eigenvector at $Re=3000$, b) zoom of a) with streamfunction isocontours superimposed for $\psi^\prime = -2.5\times10^{-4} $ to $ 2.5\times10^{-4}$ with increments for isocontours of $5\times10^{-5}$ (negative values are dashed lines), c) Same as a) for another eigenmode for $Re=3000$ with $\lambda= -0.062+0.49i$. Resolution R0 (see  Table \ref{reolutions}).}
    \label{evc2}
\end{figure}


\subsection{Non-normality of the linearised dynamics} \label{sec-nonnor}
\begin{figure}
    \centering
    \vspace{2mm}
        \includegraphics[width=0.49\columnwidth]{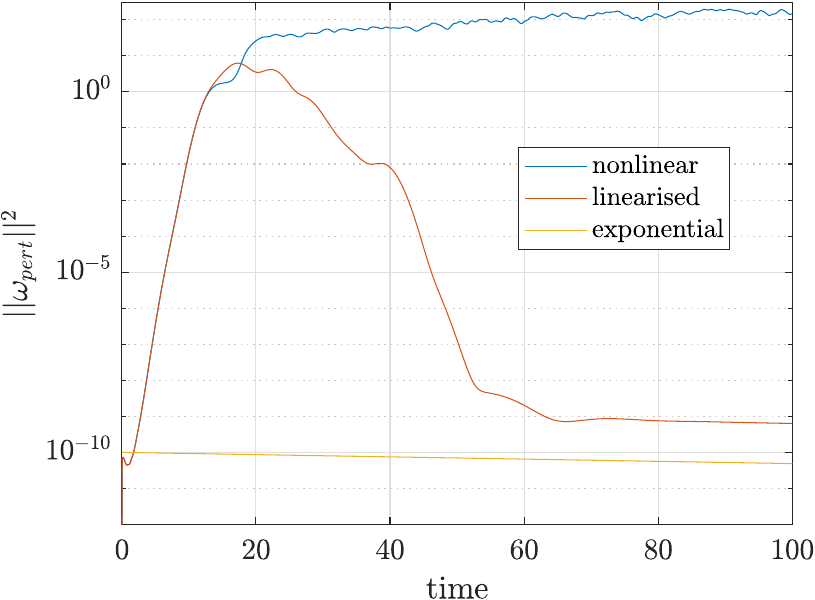}
        \includegraphics[width=0.49\columnwidth]{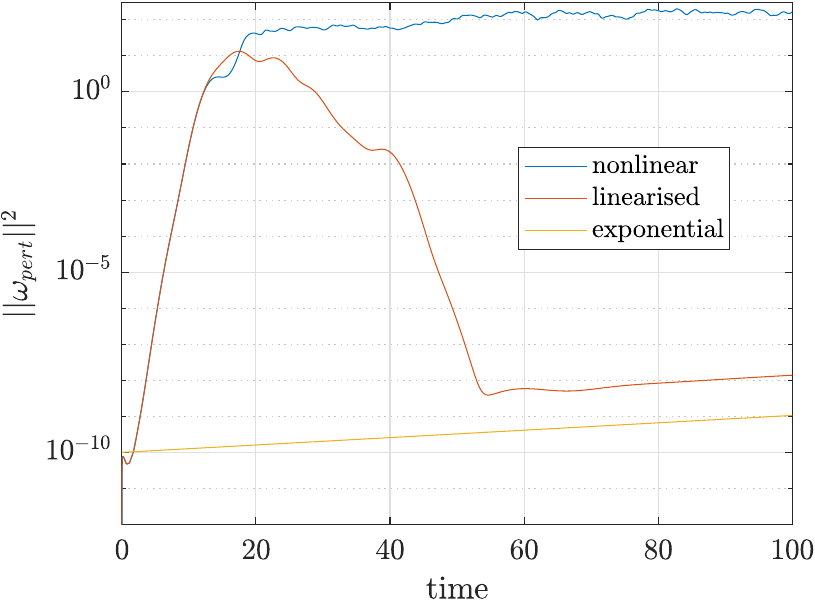}
    \caption{{Time evolution of {the observable $a^2 {(= \vert \vert \omega_{pert} \vert \vert^2)}$ in} linear and non-linear time integration at $Re=2900<Re_c$ (left) and $Re=3000>Re_c$ (right). The slope of the late time evolution of the observable in the linear code is compared with the exponential growth $\propto exp(2\lambda_rt)$ {associated with} the least stable eigenmode. {Spatial resolution R0.}} 
    }
    \label{nonnor}
\end{figure}

As noted in \cite{Daube_cf_2002} the {dynamics linearised around the base flow} in the rotor-stator cavity is strongly non-normal. To assess the strength of the non-normality a linear time integration of the Navier-Stokes equations \eref{ns} is performed. A {divergence-free}, random uniformly distributed perturbation is added {at $t=0$ to the azimuthal component ${U_b}_{\theta}$ of the base flow.} The equations linearised around the base flow are {then} advanced in time. {Rather than focusing on the traditional energy gain as in Ref. \cite{Daube_cf_2002}}, here we monitor in time the $L_2$-norm of the azimuthal vorticity perturbation 
\begin{equation} \label{obser}
        a (t)=\vert \vert \omega_{pert} \vert \vert=\sqrt{\int \vert\omega -\omega_{b}\vert^2 r dr dz},
    \end{equation} 
(where $\omega=\partial_z u_r-\partial_r u_z$ is the azimuthal vorticity component, $\omega_b$ the equivalent quantity for the base flow and $\omega_{pert}$ the difference between the two) {whose square is plotted} in Figure \ref{nonnor}.

The time { evolution} of $a(t)$ resulting from linearised and non-linear time integrations are compared in Figure~\ref{nonnor}. Initially almost identical, these signals diverge when the nonlinear code starts to {approach} the large amplitude solution branch,  a snapshot of which { has already been} shown in Figure~\ref{evc}. The observable $a^2$, initially around $10^{-10}$, is amplified by more than twelve orders of magnitude {in less than 20 time units}, before {it starts to decrease}. A growth in energy of this order was also reported in \cite{Daube_cf_2002}. High levels of non-normal growth are expected to lead to additional difficulties in the present analysis. In particular it makes it impossible to use only a time integrator to capture the base flow and one has to resort to the Newton solver instead. With a time integrator, {in the subcritical regime}, despite the formal property of linear stability of the base flow, a small perturbation {resulting from a very small increment} in $Re$ will be strongly amplified and will, \emph{in practice}, {lead to} the chaotic attractor. 

{For $Re=2900$ which is below $Re_c$, linear stability predicts a negative growth rate. This is visible here in Fig. \ref{nonnor} (left) in the asymptotic exponential decay. The quantitative negative growth rate} $\lambda=-3.58 \times 10^{-3}$ {matches well the final exponential decay (multiplied by two since $a^2$ is considered rather than $a$).} For $Re=3000$ which exceeds $Re_c$, the decrease of $a^2$ is followed by an exponentially increase. The exponential growth associated with the most unstable eigenvalue $\lambda=1.18 \times 10^{-2}$, for the corresponding $Re$, is plotted in Fig. \ref{nonnor} (right). 
The good agreement at late times confirms the prediction from linear stability analysis.



\section{Periodic solutions} \label{sec-per-sol}
Finding the critical Reynolds number $Re_c$ {associated with the first destabilisation of the laminar base flow is the} starting point in the construction of { a bifurcation diagram. 
} The rest of the article is devoted to a more complete description of this bifurcation diagram. 
Since the loss of stability {at $Re=Re_c$ is due to a pair of complex conjugate eigenvalues, it corresponds to a Hopf bifurcation, from which a nonlinear branch of oscillatory solutions is expected to arise. If the branch bifurcates towards $Re<Re_c$ and features unstable periodic orbits in the neighbourhood of $Re_c$, the bifurcation is said to be subcritical. If it bifurcates towards $Re>Re_c$ and features stable limit cycles in the neighbourhood of $Re_c$, it is said to be supercritical.
The periodic solutions can then be continued further in parameter space using arclength continuation.}


In order to identify oscillatory solutions near a Hopf bifurcation point, the recent Self-Consistent Method (SCM) \cite{mantic_prl_2014} will be used. It was previously used with success for the supercritical bifurcation of {the two-dimensional wake behind a circular cylinder} \cite{mantic_prl_2014}. A quick summary of the method in presented in Subsection \ref{sec:SCM}.

{An \emph{a priori} reasonable expectation, given the sustained subcritical states observed in Ref. \cite{Daube_cf_2002}, is that a Hopf branch of unstable solutions bifurcates subcritically directly from the base flow at $Re=Re_c$. Such a branch could ideally be continued down to low values of $Re$, before it folds back and bifurcates into new solutions that form the backbone of the chaotic attractor (as in e.g. Ref. \cite{avila2013streamwise}).} 
This expectation will be confronted with the present results in Subsection \ref{sec:Hopf}. {In the quest for the Hopf branch using SCM, another branch of periodic solutions was unexpectedly found, which will be reported in Subsection \ref{sec:per-sub-sol}. As will be shown, unlike the Hopf branch which appears {in fact} supercritical, this new branch is subcritical and does extend to low $Re$ values. }






\subsection{Self-Consistent Method (SCM) }\label{sec:SCM}
The SCM serves as {an algorithmic} tool to find the periodic solutions of the system \eref{ns} near a Hopf bifurcation point. It assumes that the periodic solution can be described as a sum of a steady field $\mathbf{U}$ {(the \emph{mean flow}, which is generally distinct from the base flow)} and an oscillatory field of complex amplitude $\mathbf{u}'$ together with its complex conjugate $\mathbf{u}'^*$ {(the \emph{oscillatory mode})}. {Denoting ${\bm U}$ the mean velocity field and $P$ the mean pressure field, the following ansatz is considered with $\lambda$ a complex number ($\lambda = \lambda_r + i \lambda_i  ) $ and $\lambda^*$ its conjugate :}
      \begin{subequations} \label{scmd}
\begin{align}
\mathbf{u}=\mathbf{U}+A(\mathbf{u}'e^{\lambda t } + \mathbf{u}'^{*} e^{{\lambda}^* t })  \\
p=P+A(p'e^{\lambda t } + p'^{*} e^{{\lambda}^* t }) 
\end{align}
\end{subequations}

      {The additional real parameter $A$ is introduced to cope with the fact that $\mathbf{u}'$ is defined up to a multiplicative constant}. 
      The ansatz \eref{scmd} is introduced into the Navier-Stokes equation \eref{ns}. {Separating the steady and oscillatory parts, and retaining only the terms oscillating at {angular} frequency {$\pm \lambda_i$}, the resulting set of equations is
      \begin{subequations} \label{scm}
\begin{align}
(\mathbf{U}\cdot \nabla)\mathbf{U} + 2A^2(\mathbf{u}'^*\cdot \nabla)\mathbf{u}' = - \nabla P + \frac{1}{Re}\nabla^2\mathbf{U}, \ \ \ \nabla \cdot \mathbf{U}=0, \\
 \lambda \mathbf{u}' +(\mathbf{u}'\cdot \nabla)\mathbf{U}+(\mathbf{U}\cdot \nabla)\mathbf{u}' = - \nabla p' + \frac{1}{Re}\nabla^2 \mathbf{u}', \ \ \ \nabla \cdot \mathbf{u'}=0.
\end{align}
  \end{subequations}

Equation \eref[a]{scm} forms a forced {nonlinear} steady Navier-Stokes equation for the field $\mathbf{U}$ with a forcing term proportional to $(\mathbf{u}'^*\cdot \nabla)\mathbf{u}'$. Equation \eref[b]{scm} results from the {formal} linearisation of the Navier-Stokes equations around the {mean flow $\mathbf{U}$}. {This amounts to neglecting all $o(A)$ terms (which is consistent with discarding terms that oscillate at {$2 \, \lambda_i$}}) even if in practice we will consider $A=O(1)$. \review{Equation \eref[b]{scm} is an eigenvalue problem of the same form as equation \eqref{evp} except that ${\bm U_b}$ is replaced by ${\bm U}$. The associated eigenvector $\mathbf{q}'=(u_r',u_{\theta}',u_z',p')$, computed using ARPACK, is normalised to unit amplitude with respect to the $\mathbf{B}$ matrix \eqref{ev} such that $\mathbf{q'}^{*}\mathbf{B}\mathbf{q}'=1$.} {By starting from suitable guesses for $\mathbf{U}$ and $\mathbf{u}'$ for a given parameter $A$, an iterative method is used to solve the two equations \eref[a]{scm} and \eref[b]{scm} alternately until convergence}. {The next value for $A$ is then sought for iteratively using a secant method until the steady state $\mathbf{U}$ becomes neutrally stable, i.e. $\lambda_r=0$.} {We emphasize that SCM operates according to two loops, an inner and an outer one. In the inner loop the equations \eref[a]{scm}-\eref[b]{scm} are solved alternately, for fixed $A$, with the result of one equation feeding the other one until convergence. 
In the outer loop the parameter $A$ is changed \review{using a secant method} until $\lambda_r=0$.} The whole procedure results in the oscillatory solution described by equations \eref[a]{scmd} and \eref[b]{scmd}. More details on the method can be found in Ref.~\cite{mantivc2015self}.
\subsection{ Harmonic Balance Method (HBM) } \label{newt-mtd}

{The Newton method discussed in section \ref{num-sim} was used to identify steady state solutions. It can be generalised for the identification of periodic orbits. While the use of a Newton method for steady states is standard procedure, its generalisation to periodic orbits has only been used recently in fluid dynamics and is thus explained in more detail here.} Exactly periodic solutions can be approximated by a finite Fourier expansion of the form

\begin{subequations} \label{newt}
\begin{align}
\mathbf{u}(\mathbf{x},t)=\mathbf{U}(\mathbf{x})+\sum_{k=1}^{nt} \{\mathbf{u}_{k}(\mathbf{x})e^{ik\omega t }+\mathbf{u}^{*}_k(\mathbf{x})e^{-ik\omega t }\} \\
{p}(\mathbf{x},t)={P}(\mathbf{x})+\sum_{k=1}^{nt} \{{p}_{k}(\mathbf{x})e^{ik\omega t }+{p}^{*}_k(\mathbf{x})e^{-ik\omega t }\}
\end{align}
\end{subequations}
where $\omega$ is a (real) angular frequency, $nt$ is a finite positive integer and the asterisk stands again for complex conjugate. The identification of periodic orbits based on this truncated Fourier expansion is sometimes described as the Harmonic Balance Method in the literature, which was recently used in different flow cases \cite{bengana_prf_2021,SierraAusin_cmame_2022}. The value of $nt$ parametrises the accuracy of the spectral approximation, with convergence guaranteed for increasing $nt$ provided the time dependence is smooth enough. 
When the periodic orbit is simple enough, for instance close to a bifurcation point, a small number of harmonics may be enough to describe it accurately, whereas for more complex periodic dynamics more harmonics will be required.

Once the finite Fourier decomposition \eref{newt} is introduced into the Navier-Stokes equation \eref{ns} as an ansatz for ${\bm u}$ and $p$, a set of nonlinear equations arises : one equation for each of the steady fields $\mathbf{U}$, $P$ and the others for each Fourier component $\mathbf{u_k}$, ${p_k}$ with $1 \le k \le nt$. This set of nonlinear equations can be solved using a Newton method to determine the fields $\mathbf{U}$, $P$, $\mathbf{u_k}$, $p_k$ for all $k$ and the unknown angular frequency $\omega$. 
A temporal phase condition is added to the list of equations to make the Newton system well-posed. \review{Indeed, the orbit is defined up to a phase shift. Different types of phase conditions can be imposed. For example, 
a value or a derivative of a variable is fixed at a specific point along the orbit. 
Imposing a derivative to zero for an arbitrary variable is generally a good choice and avoids requiring the knowledge of a specific value. Other phase conditions, less straightforward to implement, such as an integral phase condition can be imposed and is often favored when solving ordinary differential equations (see chapter 7 in \cite{seydel2009practical}). Here, as a phase condition, the time derivative of the radial velocity is fixed to zero at the beginning of the orbit and at one point in the domain.} 
The Newton algorithm is converged to machine precision {(i.e. when the Euclidean norm of the residual vector of the Newton method drops to $10^{-11}$ or less)}. Importantly, the decomposition \eref{newt} for $nt=1$ is the decomposition used in the original SCM in Ref. \cite{mantic_prl_2014} at convergence. Thus, SCM and HBM for $nt=1$ differ only in the iterative process of convergence of the solution. 
Once one solution of the Navier-Stokes equations is converged using {HBM} it can be continued in parameter space using standard arclength continuation \cite{allgower2012numerical}.

\subsection{Hopf branch \label{sec:Hopf}}

The branch of oscillatory solutions bifurcating at $Re_c$ is labelled the Hopf branch. Solutions on this branch are assumed exactly periodic, i.e. of the form \eref{newt}. When using only one oscillatory mode this form reduces to \eref{scmd}. The coupled problem \eref[a]{scm}-\eref[b]{scm} is first solved
for $A=0$ and then for $A=0.8$ ($Re=2925.5 > Re_c=2925.47$).
The vector $\mathbf{u}'$ in the forcing terms \eref[a]{scm} is selected in both cases as the 
most unstable
eigenvector
from Eq. \eref[b]{scm}. Comparing the converged spectra for $A=0$ and $A=0.8$ , the most unstable eigenvalue 
has moved towards the left upon increasing $A$ (see Figure \ref{ev1}). Therefore, a value of $A=A^*$ such that $\mathbf{U}$ becomes  neutrally stable, i.e. 
$\lambda_r = 0$, presumably exists nearby. This value $A^*$ is sought using the secant method. For the spatial resolution R0 this value is $A^* \approx 1.78$, as shown in Figure \ref{ev1}. \review{This controlled SCM procedure has proven essential in catching the Hopf branch. The basin of attraction is small, hence for a given $Re > Re_c$, the amplitude $A$ must be very close to $A^*$ so as not to diverge.} 

Finding the neutrally stable field $\mathbf{U}$ and the corresponding oscillatory field $\mathbf{u}'$ yields one solution on the Hopf branch. Continuation in parameter space allows one to progress along the branch, which is displayed in Figure \ref{h1} using the time-average scalar observable $a$ defined by \eref{obser}. 


\begin{figure}
    \centering
    \includegraphics[width=0.6\columnwidth]{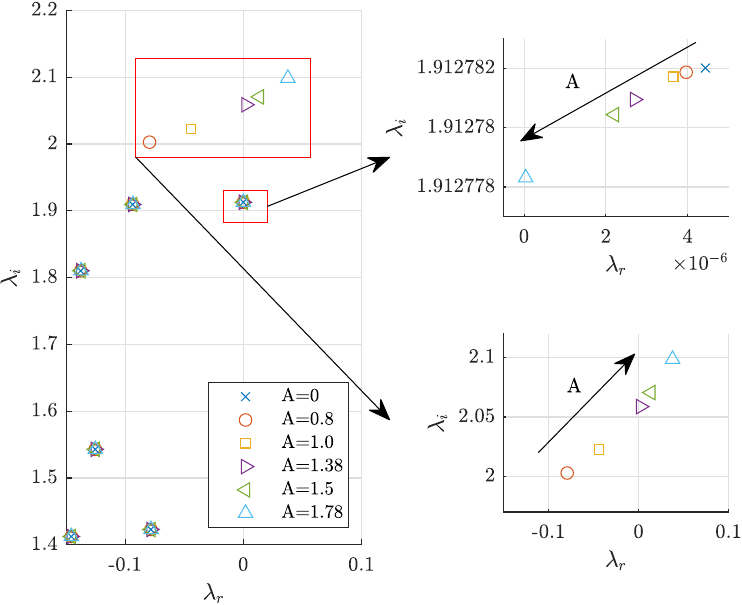}
    \caption{
    \reviewTwo{Application of self-consistent method (SCM) following Eq. \ref{scm}.}
    Spectrum evolution for $Re=2925.5 \gtrsim Re_c=2925.47$ with changing amplitude $A$ of the forcing. \review{When no forcing is applied (A=0) the base flow is unstable to a single eigenvector with the corresponding eigenvalue $\lambda=4.4\times10^{-6}+1.912782i$ (blue cross in inset).}
    \reviewTwo{Top-right inset:} eigenvalue evolution of the eigenvalue $\lambda$ leading to the Hopf branch discussed in section~\ref{sec:Hopf}. \reviewTwo{Each symbol marks the convergence of an inner loop process.  The outer loop starts at $A=0$
    and converges for $A \approx 1.78$, when the
    resulting eigenvalue has a zero real part (only a few selected values of $A$ are shown here).}
    \reviewTwo{Bottom-right inset:} evolution of another eigenvalue eventually leading to the discovery of a new branch as described in section~\ref{sec:per-sub-sol}.
    Spatial resolution R0. 
    } 
    \label{ev1}
\end{figure}

The analysis of the Hopf branch leads to a few specific observations :
    \begin{itemize}
        \item A branch of periodic solutions is born at $Re_c$ as expected
        \item {at its onset the branch bifurcates towards $Re>Re_c$}
        \item the amplitude of the oscillating solution scales like $\sqrt{Re-Re_c}$ {near the bifurcation point.}
    \end{itemize}
    
The Hopf bifurcation occurring at $Re=Re_c$ is therefore supercritical, at odds with earlier expectations. While the approximation $nt=1$ is valid close enough to $Re_c$, at a finite distance from it more temporal Fourier modes are needed, corresponding to $nt>1$.
\begin{figure}
    \centering
    \includegraphics[width=0.9\columnwidth]{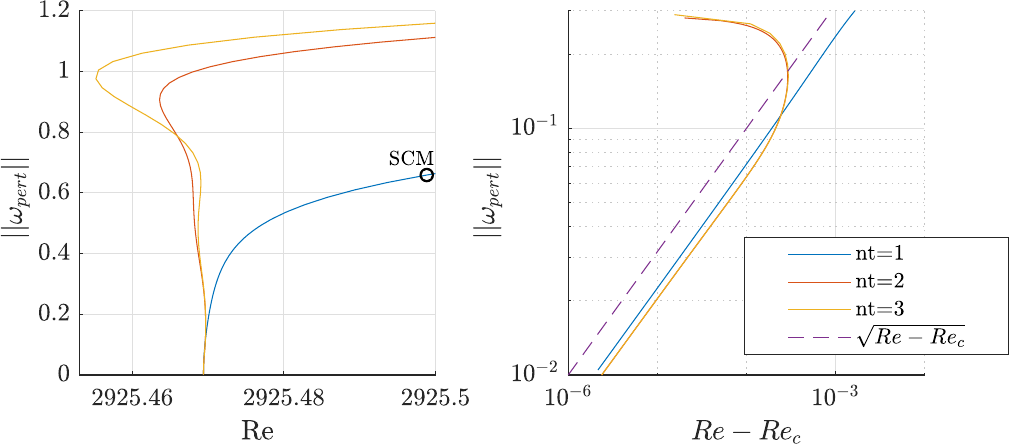}
    \caption{Left : onset of the Hopf branch for increasing parameter $nt$. The point at which the initialisation from the SCM solution {to HBM solution} is done is marked with a black circle. Right : zoom on the region $10^{-6}<Re-Re_c<10^{-3}$ with $Re_c=2925.47$, where the supercritical scaling $a \propto \sqrt{Re-Re_c}$ (dashed line) holds \reviewTwo{(logarithmic scale)}. The color scheme is the same for both plots. Spatial resolution R0. 
    }
    \label{h1}
\end{figure}
A few observations can be made:
\begin{itemize}
    \item The  solutions for $nt=1, 2$ and $3$ collapse onto each other near $Re_c$ but only {in the very narrow range {${0<Re-Re_c<10^{-4}}$.}} This means that the strong assumption of the simple form of the limit cycle is valid only very close to $Re_c$ for this flow.
    
    \item The $nt=2$ and $nt=3$ solutions collapse onto one another longer but  diverge from one another
    when $\vert \vert \omega_{pert} \vert \vert \approx0.5$. This means that {the solutions on the Hopf branch} becomes {increasingly} complicated {away from $Re_c$}. {Generally, whenever the branches for subsequent $nt$ stop overlapping they are considered temporally unresolved, and special care should be taken when interpreting the results.}
    
    \item {The Hopf branch shoots up} 
    vertically from the bifurcation point. 
    {On one hand any approximation error introduced by a finite difference scheme in time would move $Re_c$ and the branch in  parameter space, on the other hand the vanishing growth rates imply large integration times}. 
    {This explains why they were not identified in Ref.~\cite{Daube_cf_2002}. Besides the strong non-normality of the linearised dynamics around the base flow (see Subsection \ref{sec-nonnor}) carries over to the Hopf branch.} 
    The solutions on the Hopf branch are therefore believed to be virtually intractable using a time integrator. {The link between strong non-normality and the steepness of the Hopf branch near criticality is discussed for instance in \cite{Chomaz_arfm_2005}.}

\end{itemize}

The whole SCM initialisation and HBM continuation procedure {were} repeated for {the} spatial resolution R2.
The result {appears} to be qualitatively mesh-independent (not shown) and the finding of the bifurcation being supercritical is therefore robust. Recalling the expectation that a subcritical bifurcation at $Re_c$ could explain the occurrence of concentric rolls at low enough $Re\ll Re_c$, the results can appear so far as disappointing. Not only is the Hopf branch supercritical, but the several folds that occur along the branch keep this branch limited to a relatively narrow $O(1)$ range in $Re$. Although further continuation past many folds could shed more light on this question, it appears safe to claim that the Hopf branch close to $Re_c$ is probably not likely to explain the subcritical sustained states reported here and in \cite{Daube_cf_2002} at $Re\approx2000$.

{Interestingly, for $A=1.38$ another eigenvalue, characterised by $\lambda_i >2$, is close to becoming neutrally stable (see the purple triangle in the left most part of Figure~\ref{ev1}). This means that, provided the eigenvector of the forcing in the equations \eref{scm} is changed to that eigenvector, SCM can again be used {as an initial guess} to uncover 
another new branch of periodic solutions. This process is described in the following section.}


\subsection{Periodic-subcritical branch} \label{sec:per-sub-sol}


\reviewTwo{
We describe here how a new branch of periodic solutions, different from the Hopf branch, was identified numerically as a by-product of the search for the Hopf branch. As the choice of the eigenvector $\mathbf{u}'$ (resp. $\mathbf{u}'^*$) used in Eq. \ref{scm}(a) is after all arbitrary, it can be traded for any other eigensolution $\tilde{\mathbf{u}}'$  (resp. $\tilde{\mathbf{u}}'^*$) of Eq. \ref{scm}(b). $\tilde{\mathbf{u}}'$ is associated with an eigenvalue of imaginary part $\lambda_i\approx2.05$ rather than the previous eigenvalue $\lambda_i\approx1.91$. The substitution $\mathbf{u}'\leftarrow \tilde{\mathbf{u}}'$ was performed at the end of the outer iteration that resulted in $A=1.38$. The eigenvector $\tilde{\mathbf{u}}'$ chosen, associated with an amplitude denoted $\tilde{A}$, was indeed close to neutral during the resolution of Eq. \ref{scm}(a) with $A=1.38$ (see Figure \ref{ev1}), which makes it an alternative interesting guess for SCM. 
After the substitution, the mean flow is no longer neutrally stable, but the SCM
inner and outer iterations can resume.
During this second process we observe a substantial increase in $\lambda_i$ from $\approx2.05$ to $\approx2.3$. At convergence the new amplitude 
$\tilde{A} \approx 1.48$. 
 The first and the last step of the secant method (the outer loop) are illustrated in Figure \ref{fig:scm-evol}. }

}

}

\begin{figure}
    \centering
    \includegraphics[width=0.5\textwidth]{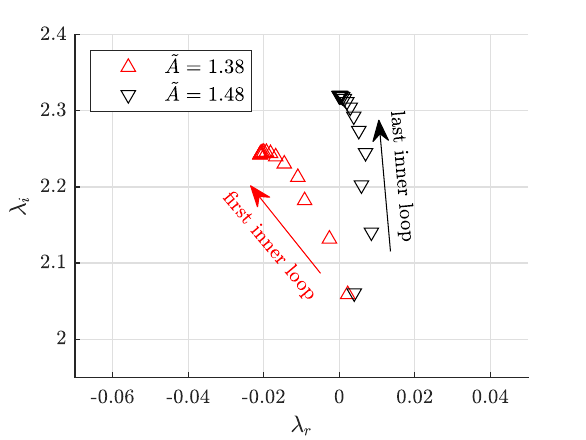}
    \caption{
    \label{fig:scm-evol}
    \review{\reviewTwo{Determination of the periodic-subcritical branch using SCM.} \reviewTwo{
    At $A=1.38$ the eigenvector ${\bm u'}$ was changed to another eigenvector ${\bm \tilde{\bm u}'}$ (see text) with an amplitude denoted as $\tilde{A}$. $\tilde{A}$ is set to $1.38$ for the first inner loop. The red triangles show the convergence of the inner loop for $\tilde{A}=1.38$ with the final $\lambda_r\neq0$.
    After achieving convergence on $\tilde{A}$ the last inner loop is shown using black triangles for $\tilde{A}=1.48$ with the final $\lambda_r\approx0$ and corresponding $\lambda_i\approx2.33$. This convergence process on $\tilde{A}$ takes approximately 10 iterations. For the sake of clarity only the first and the last inner loops are shown. 
    } 
    }}
\end{figure}

\begin{figure}
    \centering
   \includegraphics[width=0.7\columnwidth]{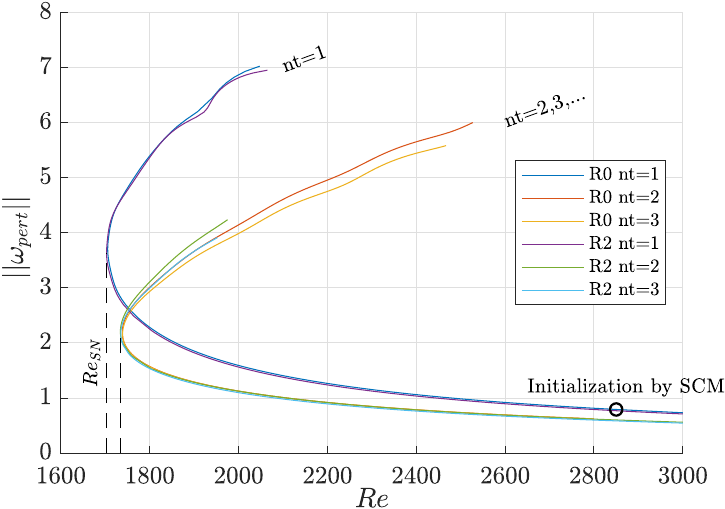}
    \caption{Periodic-subcritical \reviewTwo{branch} for discretisations R0 and R2 \reviewTwo{and increasing $nt$, where} $nt$ is the number of Fourier modes in the decomposition \eref{newt}. The point at which the initialisation from the SCM solution {to HBM solution} is done is marked with a black circle. Branches fold around $Re=Re_{SN}\approx1700$. The exact saddle-node bifurcation values are listed in Table \ref{resn}.}
    \label{fig:g1}
\end{figure}

 When SCM is sufficiently converged, the solution is used to initialise HBM. HBM is used \reviewTwo{to converge the solution, which is} then continued in parameter space (see Figure \ref{fig:g1}).

The above methodology yields a 
branch 
very different from the Hopf branch.
This branch extends down to $Re \approx 1700$, i.e. much lower than both $Re_c$ and the Hopf branch altogether. The newly found branch is hence called the \emph{periodic-subcritical branch}. \reviewTwo{Precise values of $Re$ at which this saddle-node bifurcation happens, $Re_{SN}$, are documented in Table \ref{resn}.} Its lower part also extends towards values of $Re \gg Re_c$, at least {up to} 7000 for both grid resolutions (not shown). This branch was therefore \emph{not} found to bifurcate from the steady state in the range of $Re$ investigated. The angular frequencies of the solutions on this branch are reported in Table \ref{angularfreq}.\\

\begin{table}[]
\begin{tabular}{c|ccc}
nt & 1 & 2 & 3 \\ \hline 
R0 & 1705.3 & 1738.1 & 1738.9 \\
R2 & 1702.8 & 1733.9 & 1734.7
\end{tabular}
\caption{Saddle-node value $Re_{SN}$ depending on the spatial and temporal discretisation.}
\label{resn}
\end{table}

\begin{table}[]
    \centering
    \begin{tabular}{l|lllllll}
$Re$ & 1900   & 1800   & \textbf{1734.7} & 2000   & 2300   & 2600  & 3000   \\ \hline
$\omega_{per-sub}$ & 3.934 & 3.586 & \textbf{3.136} & 2.559 & 2.421 & 2.351 & 2.297 \\
$f$ & 0.626 & 0.570 & \textbf{0.499} & 0.407 & 0.385 & 0.374 & 0.365
\end{tabular}
    \caption{Angular frequency of the periodic solutions on the periodic-subcritical branch shown in Figure \ref{fig:g1}. The point corresponding to the saddle node bifurcation is emphasised in bold. $Re$ to the left (right) of this point correspond to the top (bottom) part of the periodic-subcritical branch in Figure \ref{fig:g1}. $\omega_{per-sub}$ corresponds directly to the angular frequency used in formula \eref{newt}. Resolution R2, $nt=3$. }
    \label{angularfreq}
\end{table}

\begin{figure*}
    \centering
    \includegraphics[width=\textwidth]{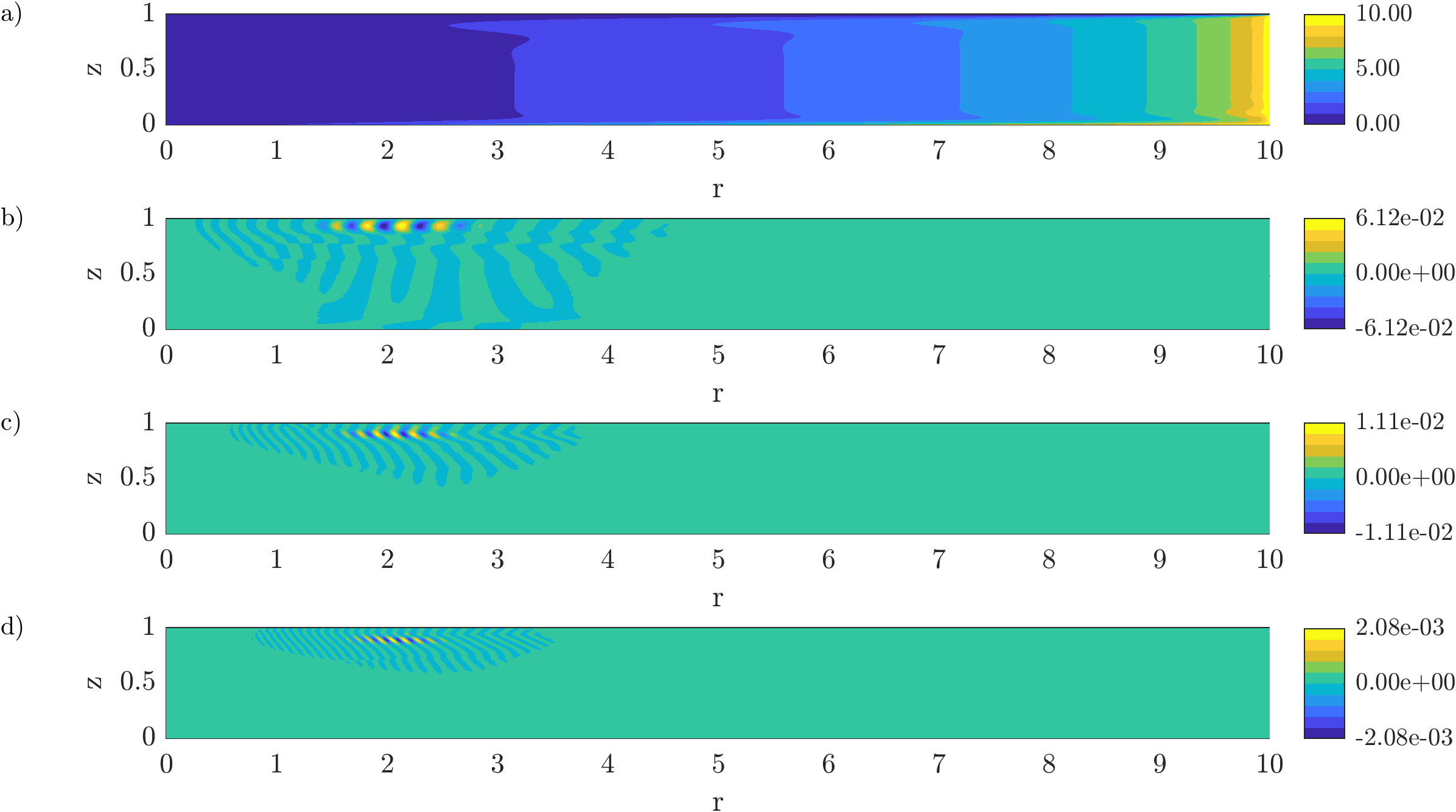}
    \caption{a) Contours of $U_{\theta}$. b-d) $u_{\theta}$ for temporal harmonics  $k=1,2,3$ (from top to bottom) for the lower periodic-subcritical solution at $Re=2300$. Spatial resolution R0, $nt=3$.}
    \label{per-sub}
\end{figure*}

{Since the periodic-subcritical branch exists over a wider range of $Re$ values than the Hopf branch, the corresponding solutions are studied in more detail.} In particular a solution from the {lower} branch is visualised in Figure \ref{per-sub} for $Re= 2300$. {It takes the form of rolls in the Bödewadt layer visually comparable to the most unstable eigenvector for the $Re \approx Re_c$}. {The decomposition in Eqs. \eref{newt}, used in the Newton method, allows for a direct analysis of the temporal harmonics of this solution}. Subsequent oscillation modes are characterised by increasingly smaller structures. {Any improvement in the temporal resolution $nt$ must hence be accompanied by an increase in spatial resolution ($N_r \times N_z$).} 


\section{Edge states} \label{sec-bis-branch}

{In this section, we focus on another type of finite-amplitude solution likely to contribute to the bifurcation diagram of the system. The third branch of solutions reported in the present work is {a} branch of \emph{edge states}. As we shall see, their dynamics is more complex than that of the periodic states of Subsection \ref{sec-per-sol}. They are known to play a major role as mediators between the base flow and {the chaotic state  \cite{khapko2016edge}, here the chaotic rolls}.}

\subsection{Notion of edge state \label{sec:edge}} 

 Edge states are unstable finite-amplitude states, generally associated with low levels of perturbation energy, that are specific to the subcritical regime where the base flow is linearly stable. They have the defining property that their instability leads both to the chaotic solutions or to the base flow, depending on the perturbation considered~\cite{skufca2006edge}. Mathematically, in the state space associated with Eq.~\eref{ns}, edge states are local attractors belonging to the intersection of the respective attraction basins of attraction of the base flow and of the chaotic set. They exist as soon as the base flow has a non-trivial basin of attraction, i.e. in  subcritical conditions, here for $Re<Re_c$. The intersection of the boundaries of these basins of attraction is the stable manifold of the edge state, sometimes called simply the \textit{edge}. The first computation of edge tracking in fluid flows is due to \cite{itano2001dynamics} in channel flow. Later computations have shown that edge states are generally spatially localised structures \cite{duguet2009localized,khapko2016edge}. Importantly, the dynamical nature of an edge state is not known in advance, it can be a steady state, a periodic orbit, or a more complicated object such as a chaotic set \cite{wang2007lower,duguet2008transition,khapko2014complexity}. More details on {the concept of edge state} can be found in \cite{skufca2006edge,schneider2007turbulence,duguet2008transition,duguet2009localized,khapko2013localized}. \review{The technique is particularly useful when combined with symmetry
subspace restrictions \cite{duguet2008transition,lopez2017transition}.} The aim of this section is to report, for the first time in the literature, calculations of edge states in rotor-stator flow for a set of Reynolds numbers. 


\subsection{Bisection algorithm \label{sec:bisection}}

{Following the above definition, the main idea behind the bisection algorithm is to find two points bracketing the stable manifold sufficiently close, such that the associated trajectories approach transiently the edge state.
Starting with two flow states, one labelled \emph{turbulent} and the other \emph{laminar}, a new initial condition is formed using the interpolation}
\begin{equation}\label{albis}
    \mathbf{u}(t=0)=(1-\alpha)\mathbf{u}_{laminar}+\alpha\mathbf{u}_{turbulent},~0<\alpha<1.
\end{equation}
This linear combination of the laminar state $u_{laminar}$ and a turbulent state $u_{turbulent}$ is the new initial condition for time integration. Depending on whether this new initial condition evolves towards the turbulent or laminar state, the initial condition is labelled as turbulent or laminar for the next iteration. This yields a sequence of $\alpha$ values $\alpha_0$, $\alpha_1$,... The process is repeated until the sequence of $\alpha_k$'s has converged. {After each iteration the bracketing interval is halved, therefore the value of $\alpha$ for which the simulation shadows the edge the longest is determined up to machine precision (taken to be $10^{-16}\approx2^{-53}$) after approximately $|log(10^{-16})/log2| \approx$ 53 iterations.} {This apparent limitation} of the bisection algorithm {by finite machine precision} can be easily overcome by a \emph{restart procedure} \cite{skufca2006edge}. {By restarting the bisection from a later time it can run indefinitely and the edge state is  reached asymptotically.}


{The same observable $a(t)$ defined in Eq.~\eref{obser} is  used to monitor edge trajectories and the bracketing trajectories. If $a (t \rightarrow \infty ) = 0 $ the trajectory has reached the laminar base flow. 

\subsection{Edge branch}
The bisection procedure is now used to identify edge states in the rotor-stator flow. {The time history of the observable $a(t)$, the vorticity perturbation norm, is reported for $Re=2300<Re_c$ in  Figure \ref{bis}. Some guiding values of the interpolation coefficient $\alpha$ are provided in the legend. Bisection successfully identifies a state that is unstable and lies, by definition, at the boundary of the basins of attraction of the chaotic and the laminar regime.}
\begin{figure}
    \centering
        \vspace{2mm}
    \includegraphics[width=0.6\columnwidth]{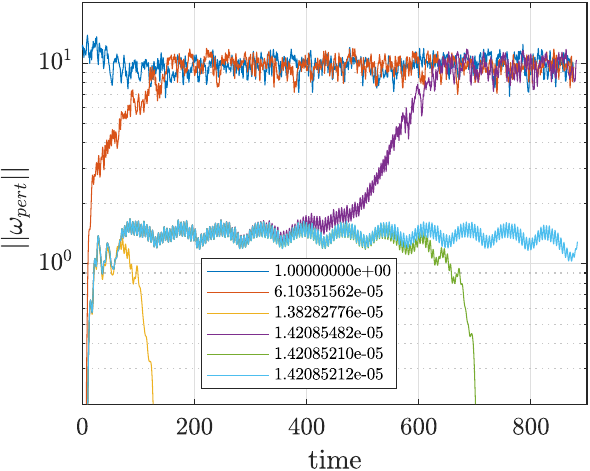}
    \caption{Time series of the observable $a(t)$ during the bisection procedure for Re=2300. Legend : coefficient $\alpha$ (defined in equation \eref{bis}) parametrizing the initial condition. Spatial resolution R0.}
    \label{bis}
\end{figure}
The same bisection procedure is repeated {from $Re \approx Re_c$ down to where bisection fails due to the apparent lack of chaotic attractor.} A representative portion of the corresponding {time series} {is displayed in} Figure \ref{edgeRe}. The signal on the edge seems chaotic for $Re=1850$, almost periodic for $Re=2000$ and apparently bi-periodic for $Re>2300$. We observe that the bisection algorithm detects also the unstable solution at $Re=3000$ which is only a few $\%$ above $Re_c=2925.47$. Since $Re>Re_c$ this state is not formally an edge state because {the laminar basin has collapsed to a single point. As demonstrated in Ref. \cite{beneitez2020modeling} using a low-order model, the edge state can still exist as finite-amplitude solution of Eq. \eref{ns} beyond $Re_c$ although it loses its property of state space mediator. This collapse of the edge state makes it however, in principle, undetectable by standard bisection. Here the reason is simpler and due to the finite time over which the edge has been followed. Since the base state at $Re=3000$ is only slightly unstable, it takes much more time for the unstable perturbation to the base flow to grow than it takes for the perturbation to the unstable edge state to grow. This explains why the bisection algorithm still detects an unstable solution near $Re_c$ even for $Re \gtrsim Re_c$.} The edge state found at $Re=2300$ is chosen for further analysis because of its simple dynamics and its apparent bi-periodicity, \review{an original property reported only lately in Refs. \cite{lopez2017transition} and \cite{bengana2019bifurcation} in the context of lid-driven cavities}.

\begin{figure}
    \centering
        \vspace{2mm}
    \includegraphics[width=\columnwidth]{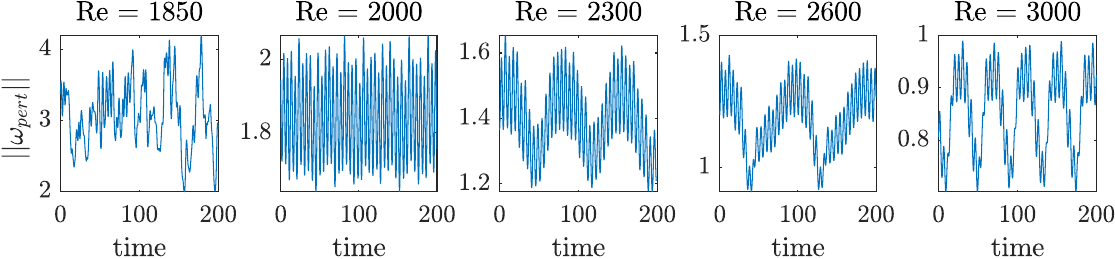}
    \caption{{Perturbation vorticity signal $a(t)$ corresponding to the edge state, for the same values of $Re$. {From left to right, $Re=1850$, $2000$, $2300$, $2600$ and $3000>Re_c$}.}}
    \label{edgeRe}
\end{figure}

\subsection{Frequency analysis}

{A first glance at the observable time series for $Re=2300$ suggests} the presence of two {main} frequencies in the edge state: one corresponding to the long period $T\approx75$, and one corresponding to a {shorter} period $T\approx5$. {For a first identification of} the interesting frequencies the {Fourier transformed of the \emph{global} observable $a(t)$ is used.} It is preferred {to the velocity signal from a \emph{local} probe because, as a global quantity, it gives an overview of {\it all} frequencies contributing to the edge state.} A part of the signal corresponding to the final bisection iteration seen in Figure \ref{bis} is extracted and its frequency spectrum is {computed}.  Care has been taken to {disregard} the initial and final transients of the signal. {The remaining time interval available for the Fourier transform} contains only approximately 6 long periods, hence finite-time effects should be expected to pollute the subsequent analysis. The frequency spectrum {$\hat{a}(f)$} in plotted in Figure \ref{edgeFreq} {as a function of the frequency $f$.} { A number of distinct discrete peaks can be observed.} {The} frequencies marked with red circles are listed in increasing order in Table \ref{freqs}. {The slow frequency ($f=0.013$) in the spectrum corresponds to the long period} seen in the observable signal ($T\approx75$). A series of peaks corresponding to $f=0.197,0.392,0.590$ {correspond to the fast oscillations of $a$}. 

\begin{figure}
    \centering
        \vspace{2mm}
        \includegraphics[width=0.8\columnwidth]{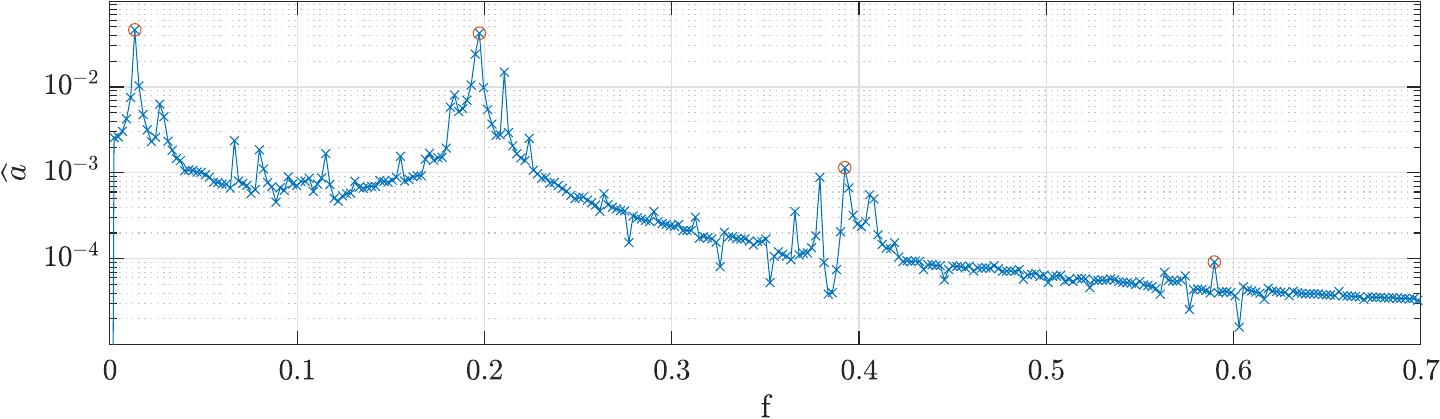}
        
    \caption{Frequency spectrum $\hat{a}(f)$ of the observable $a$. {Spatial resolution R0. The} spectrum is calculated using 2043 samples of the signal {sampled} uniformly {in the time interval \review{$t\in(300,750)$}. {The maximum frequency computed is $f_{max}=2.27$ with a frequency step $df=0.0022$}.}  The frequencies marked with red dots are listed in Table \ref{freqs}.}
    \label{edgeFreq}
\end{figure}
\begin{table}[]
\begin{tabular}{l|llll}
Frequency & 0.013  &  0.197  &  0.392  &  0.590 \\
 \hline
Period & 75.175  &  5.068  &  2.548  &  1.695 \\
\end{tabular}
\caption{{Frequencies and periods of the main peaks from the frequency spectrum $\hat{a}(f)$ in Figure \ref{edgeFreq}.}}
\label{freqs}
\end{table}

Further {insight} is gained by {computing} the Fourier transform of the {(discrete)} velocity field itself. {For this purpose a series of flow fields is saved. A snapshot of the associated perturbation velocity is shown in Figure \ref{edgeSnap} {together with a space-time diagram. The space-time diagram shows clearly the pairing and merging of the vortical structures as they propagate towards the axis inside the envelope (between $r=1$ and $r \approx 3.5$). Interestingly, whereas we are now describing an edge state which by definition is unstable, the same pairing phenomenon has also been reported, for a smaller aspect ratio, as a feature of the propagation of the rolls observable experimentally \cite{Schouveiler_jfm_2001}}. A Fourier transform is {computed} at each point of the (discretised) domain using 2043 {snapshots sampled uniformly} over the time {interval} $t\in(0,450)$. 
After bandpass filters around frequencies $f=0.013,0.197,0.392$ and $0.590$ are applied}, the velocity field corresponding to each {frequency is} reconstructed in physical space. {The azimuthal velocity components associated with each  frequency are shown in Figure \ref{edgeAn}. \\
 
 
 

 The {velocity} fields corresponding to higher frequencies ($f=0.197,0.392$ and $0.590$, Figure \ref{edgeAn}(b-d)) feature a {wavetrain of counter-rotating vortices located inside} the Bödewadt layer. {These waves have} negative radial phase velocity, i.e. they {propagate} in the direction of decreasing radius, {as is clear from} the spacetime diagram in Figure \ref{edgeSt} (bottom). Interestingly, the waveform corresponding to $f=0.392$ bears a high resemblance to the $k=1$ mode of the periodic subcritical solution reported in section \ref{sec:per-sub-sol}, particularly in Figure \ref{per-sub}. For $Re=2300$, the frequency of the periodic subcritical solution is $f_{per-sub}=0.3854$, whereas the {corresponding} frequency for the edge state is $f_{FT}=0.3924$.\\

The {Fourier mode at the} slow frequency ($f=0.013$, Figure \ref{edgeAn}a) {features vortical} structures {{at the edge} of the B\"odewadt layer for} $r\in(1,3)$ and $z\in(0.8,1)$. {Using a space-time diagram} {computed for $z=0.5$,} an apparent wavepacket of nearly vanishing radial {group and phase velocity is prominent for} $r\in(1,3)$ and $z\in(0,0.8)$, see Figure \ref{edgeSt} (top). {Vertical structures {exist also outside the B\"odewadt layer}. Consistently with the dispersion relation of inertial waves in flows with solid body rotation, such structures can be interpreted as low-frequency inertial waves.} These slow {waves, interpreted as} spatially localised standing waves (with vanishing phase velocity), are also reminiscent of the {internal} waves observed {inside the bulk of differentially heated} cavities  \cite{thorpe1968standing,oteski2015quasiperiodic}. \\

{As a yet stronger sign that the periodic and the edge solutions are connected, the Fourier mode extracted from the edge state corresponding to $f=0.392$ is used as an initial guess for HBM with $nt=2$. This is achieved in practice by building an initial guess for Newton's with $(\bm{U},P)$ as the mean solution, the edge mode corresponding to $f=0.392$ as the first Fourier mode and the corresponding angular frequency.
Convergence {towards the formerly found periodic solution} is obtained for $nt=2$ in 15 Newton iterations.} 
 This {indicates} that the SCM method was \emph{a posteriori} not {the unique way } to {determine numerically} the periodic-subcritical branch. 
 The almost identical frequencies and waveforms suggest that the periodic-subcritical solution found in section \ref{sec:per-sub-sol} is itself an unstable solution embedded in a less unstable edge state. 
 This situation happens to be common when edge states are chaotic \cite{duguet2008transition}, but deserves closer inspection here where the edge state is quasiperiodic in time. {A simple possibility is that the edge state bifurcates, directly or indirectly, from the subcritical periodic branch. This} is left for future investigation.\\

\begin{figure}
    \centering
    \includegraphics[width=1\columnwidth]{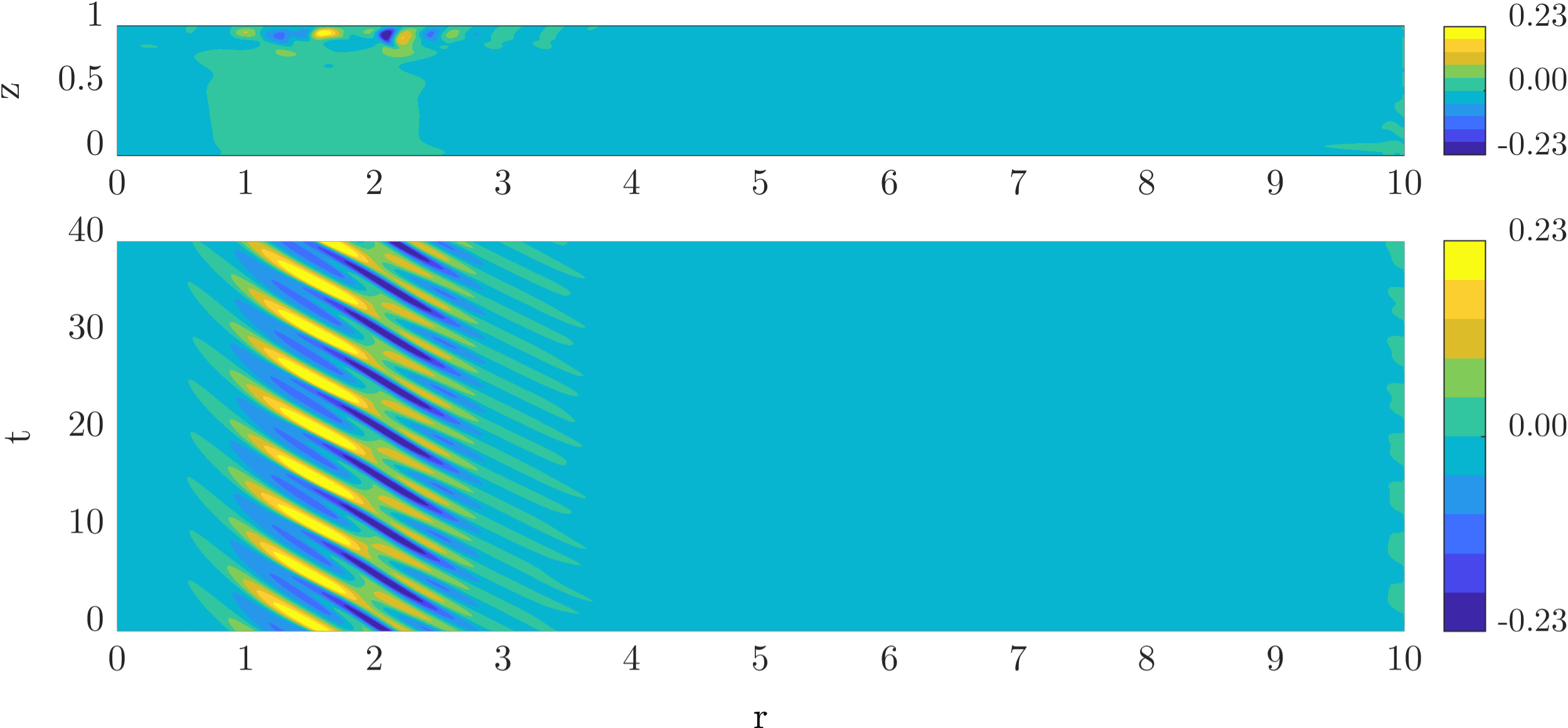}
    \caption{Top : Edge state snapshot of $u_{\theta}(r,z)$ in perturbation mode. Bottom : $(r,t)$ space-time diagram of $u_{\theta}(r,z)$ at $z=0.9375$ (bottom) for $Re=2300$. Rolls are localised in the B\"odewadt layer and travel towards the axis. Spatial resolution R0. A video that shows the temporal evolution of $u_r$ and $u_{\theta}$ in the meridional plane, during the instability of the edge state, is available as Supplemental Material (see the first half of the movie)\cite{sup_mat_1}
    . 
    }
    \label{edgeSnap}
\end{figure}
\begin{figure}
    \centering
            \raisebox{1in}{a)}
    \includegraphics[width=0.97\columnwidth]{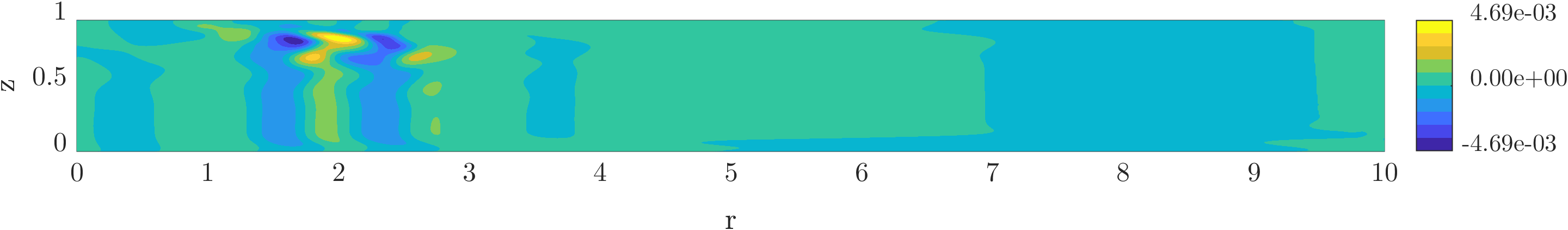}
    \raisebox{1in}{b)}
    \includegraphics[width=0.97\columnwidth]{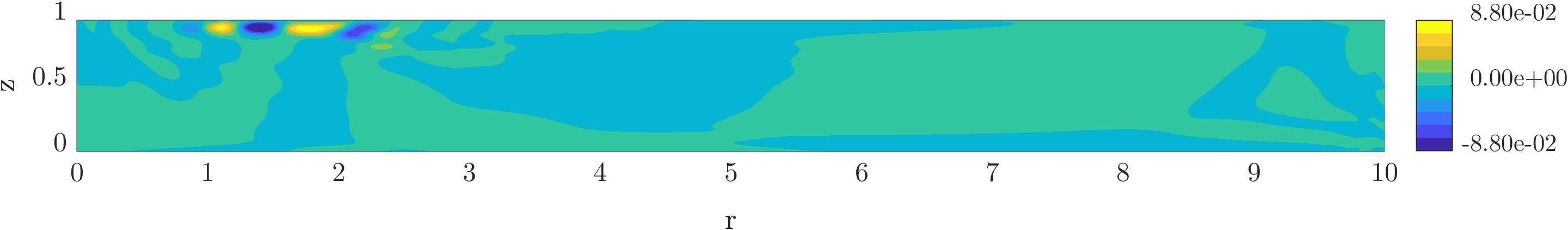}
    \raisebox{1in}{c)}
    \includegraphics[width=0.97\columnwidth]{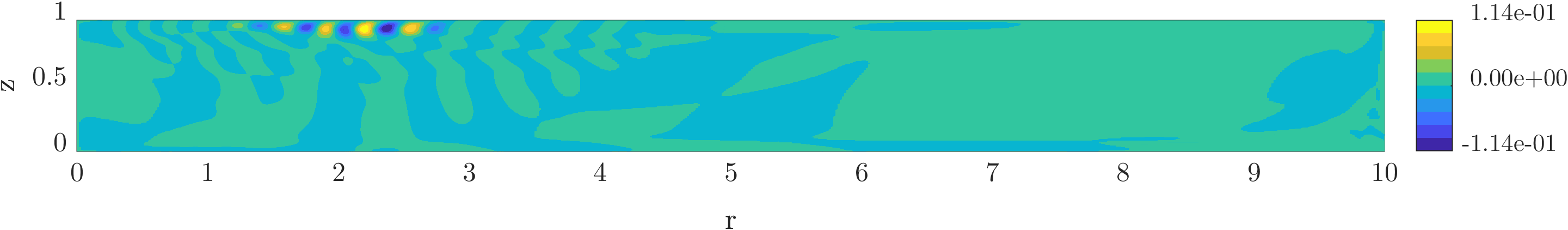}
    \raisebox{1in}{d)}
    \includegraphics[width=0.97\columnwidth]{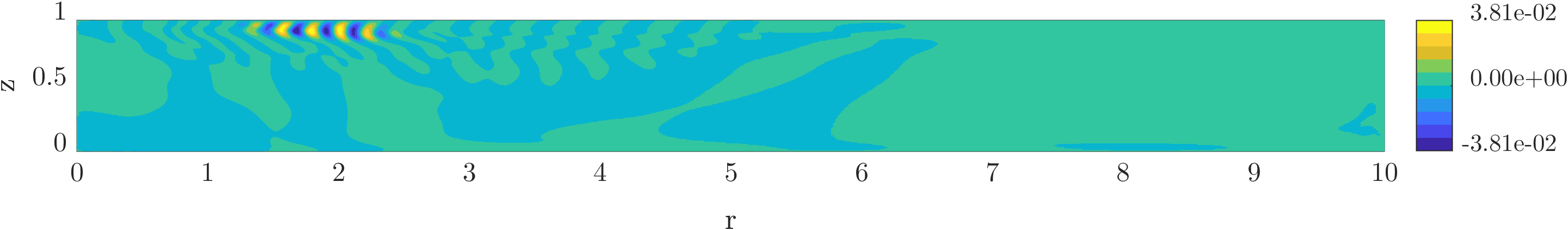}
    \caption{
    Temporal Fourier components of the edge state corresponding to  frequencies a) $f=0.013$, b) $f=0.197$, c) $f=0.392$ and d) $f=0.590$, as shown in the spectrum in Figure \ref{edgeFreq}. Associated movie available in the Supplemental Material \cite{sup_mat_2}
    . Note the resemblance between panel c) and the $k=1$ harmonic in Figure \ref{per-sub}.
    }
    \label{edgeAn}
\end{figure}
\begin{figure}
    \centering
        \vspace{2mm}
      \includegraphics[width=1\columnwidth]{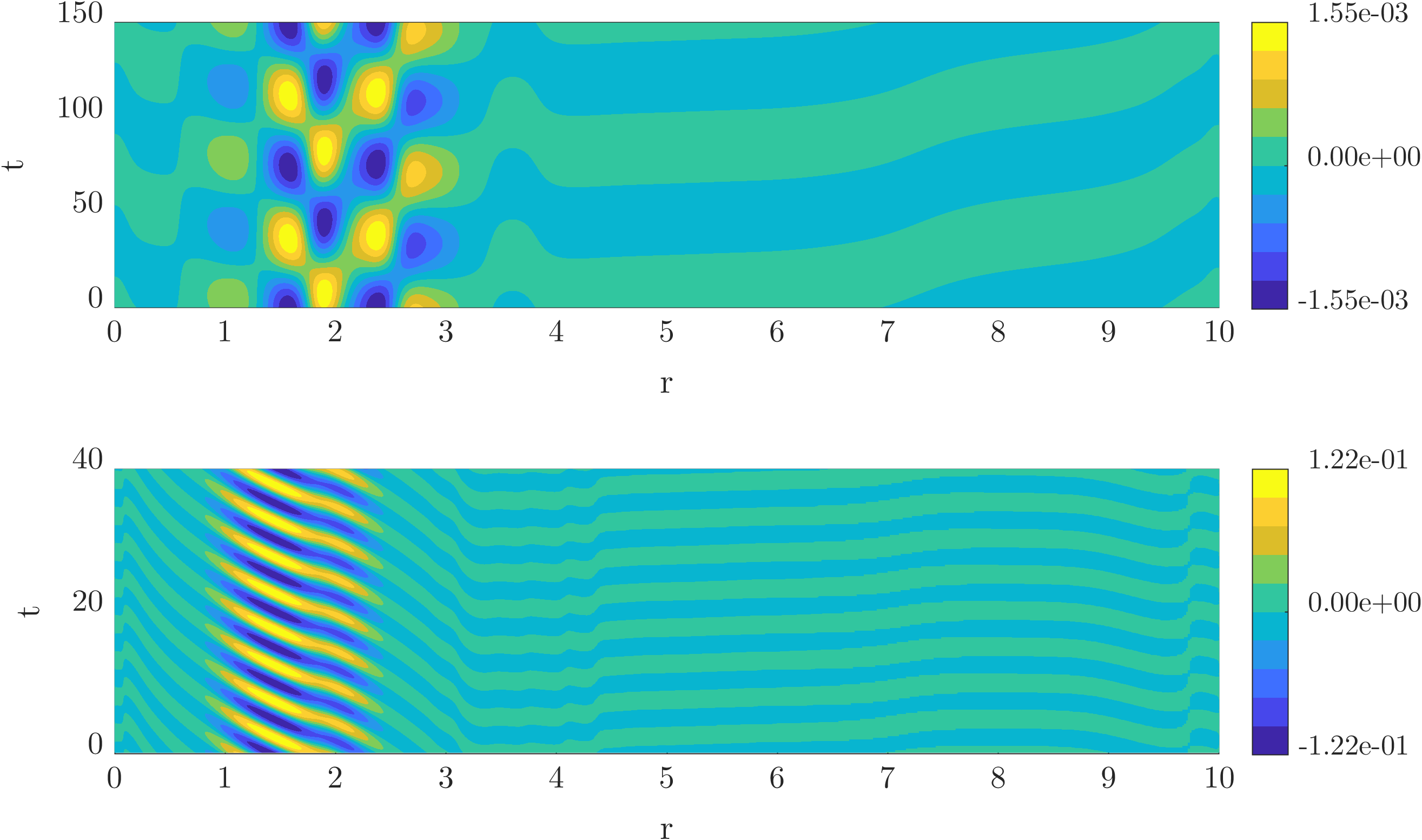}
    \caption{$(r,t)$ space-time diagrams of selected temporal Fourier modes for the edge state at Re=2300. Spatial Resolution R0. Top: mode $f=0.013$, $z=0.5$ (mid-plane cut, compare with Figure \ref{edgeAn} a), bottom: mode $f=0.197$, $z=0.9375$ (B\"odewadt layer cut, as in Figure \ref{edgeAn} b). The timescale and amplitude scale differ between the two figures. }
    \label{edgeSt}
\end{figure}

 \begin{figure}
     \centering
     \includegraphics[width=\textwidth]{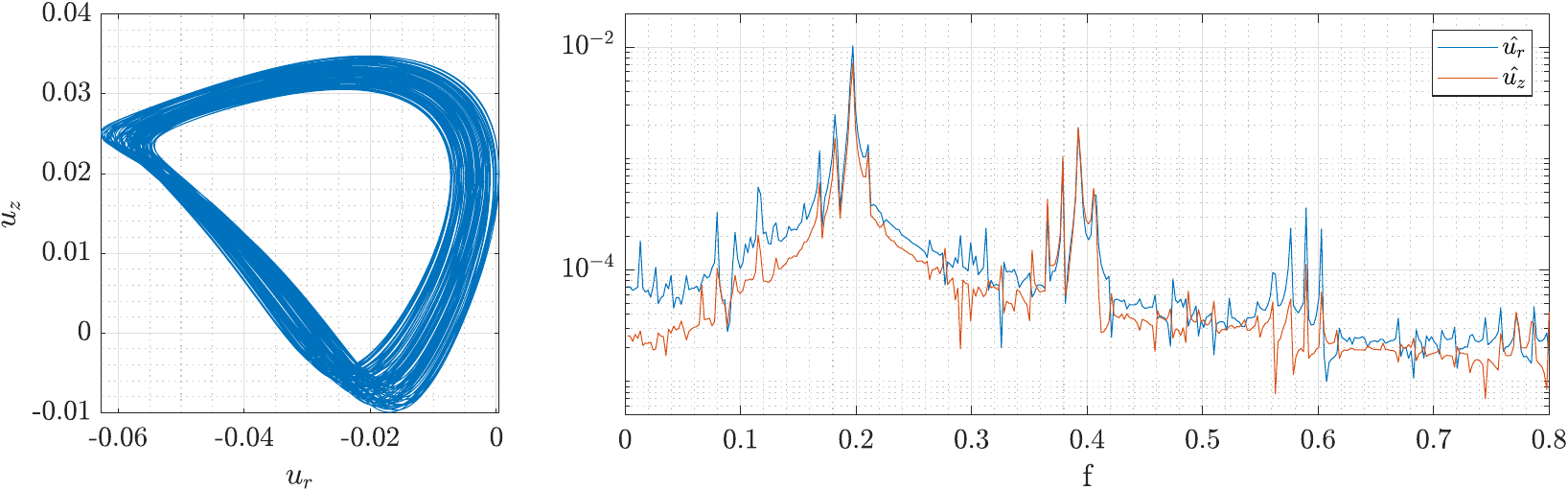}
     \caption{Velocity data at position \review{$r=1.0$}, $z=0.9375$ for the edge state at $Re=2300$. Left : phase portrait using in-plane velocities $u_r$ and $u_z$ and the spectrum of probe signal. Right : Frequency spectrum computed using $\approx100k$ samples spaced uniformly in $t\in(0,450)$. Maximum frequency : $f_{max}=113$ with a resolution $df = 0.0022$. Spatial resolution R0. }
     \label{pasesp}
 \end{figure}
 
\subsection{Quasiperiodicity of the edge state} \label{quasi:edge}

 In order to validate the hypothesis of a quasiperiodic edge state for $Re=2300$, a velocity probe is considered where the rolls achieve large amplitudes in Figure \ref{edgeSnap}. A state portrait based on the in-plane velocity components {$u_r$ and $u_z$} is shown in  Figure \ref{pasesp} (left). The plot on the right displays
 the Fourier transforms of $u_r(t)$ and $u_z(t)$.
 {The Fourier transform displays clear distinct peaks over a non-zero background. It was checked,
  by artificially shortening the time series, that the background level is not the signature of a chaotic signal, instead it is due to the finiteness of the signal (which is an inherent limitation of the bisection technique).} The {main} peaks correspond to the frequency $f_1=0.197$ and its harmonics, and to $f_2=0.013$ and its harmonics, consistent with the spectrum of $a(t)$ shown in Figure \ref{edgeFreq}. The {quadratic} interplay of these two main frequencies {explains} the other frequencies $f=n\, f_1\pm m \,f_2$ with $(n,m)\in \mathbb{(N \times N)}$   visible in the spectrum. Due to $f_1$ and $f_2$ being far apart in the discrete spectrum it is postulated the edge state of the studied rotor-stator flow at the $Re=2300$ is indeed generically biperiodic with two incommensurate frequencies. {In other words} it forms a 2-torus in the {state space of Eq. \eref{ns}}. {This property holds in the interval for most values of $Re$.
  No periodic edge state was identified for the $Re$ studied.

\subsection{{Connection between the edge state and the chaotic rolls}} \label{instab:edge}

 \begin{figure}
     \centering
                    \includegraphics[width=\textwidth]{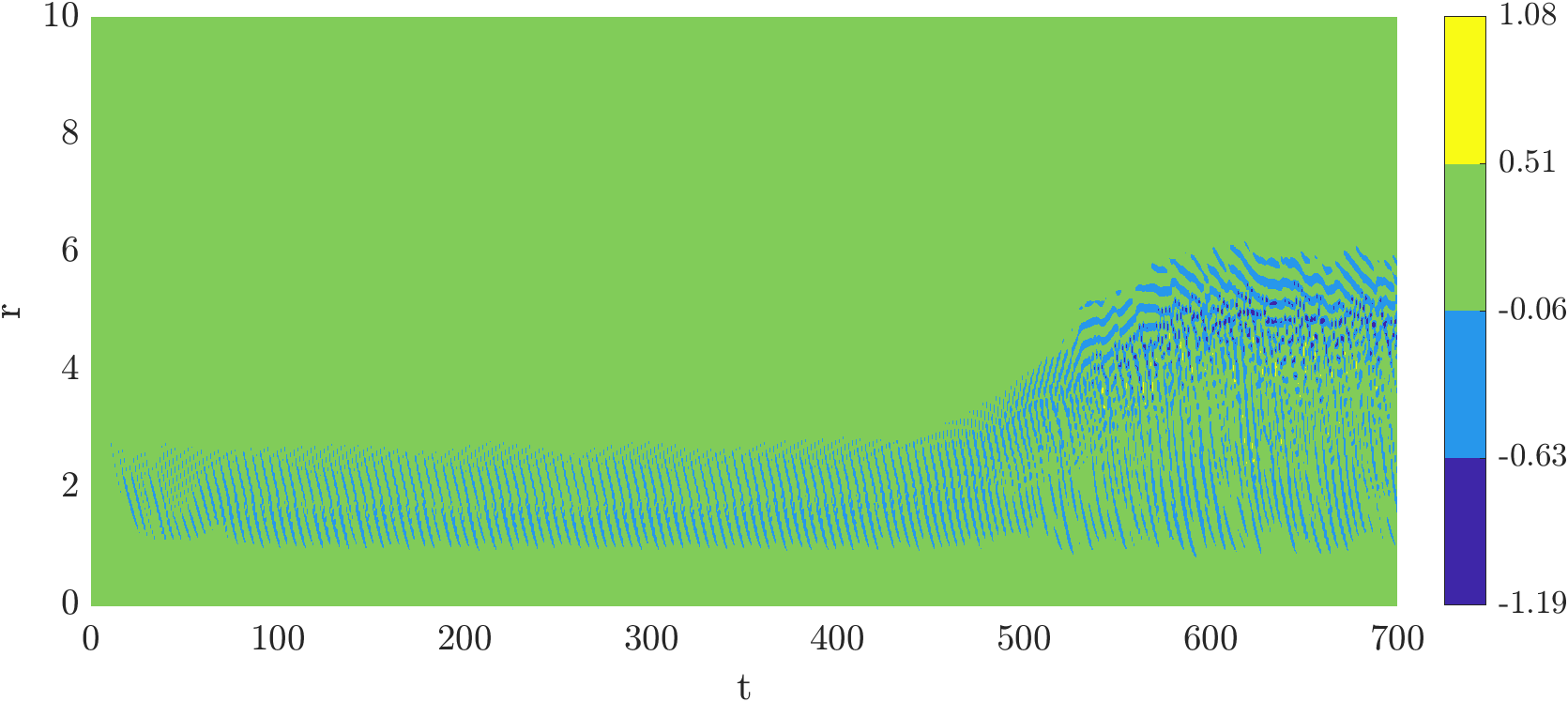}
     \caption{Instability of the edge state. $(r,t)$ Space-time diagram for the  perturbation $u_{\theta}(r,t)$ at $z=0.9375$ for $Re=2300$. The time series corresponds to one of the {\emph{non-converged}} trajectories computed  during the bisection process (corresponding to the value $\alpha=1.42085482 \times 10^{-5}$, see Fig. \ref{bis}). The departure of the edge state becomes evident for $t \ge 400$. The chaotic roll state is reached for $t \ge 600$. A movie that shows the temporal evolution of $u_r$ and $u_{\theta}$ in the meridional plane is available as Supplemental Material \cite{sup_mat_1} 
     . }
     \label{fig:edge2rolls}
 \end{figure}

 {By construction, the edge state is a linearly unstable solution
 of Eq. \eref{ns} with a specific property : typical perturbations lead, depending on their sign, either to the laminar base flow or to another chaotic state. We exploit this property and report now the dynamical path from the edge state to the attracting chaotic state \cite{duguet2010slug}. This is achieved simply by selecting and analysing one of the trajectories displayed in Fig. \ref{bis} for $Re=2300$. A space-time diagram of the azimuthal velocity perturbation along the line $z=0.9375$ is shown in Fig. \ref{fig:edge2rolls}. The dynamics follows closely the biperiodic dynamics of the edge state
until $t \approx 400$. From $t \approx 400$ to $t \approx 600$, the perturbation to the base flow undergoes a rapid exponential growth in both energy and observable $a$ (see Fig. \ref{bis}). For $t \gtrsim 600$ the stationary chaotic state is reached, with no sign of convergence to any simpler flow state. During the exponential growth phase the radial extent of the edge state grows with time. Whereas the low-$r$ end stays constant to $\approx 1$, the large-$r$ of the localised structure moves upstream from $r \approx 2.5$ to $r \gtrsim 5$. This modification results both from an intensification of the perturbations convected away from the corner/shroud region, and from the advection of perturbations \emph{against} the B\"odewadt layer (the positive phase speed visible from $t \approx 500$ to $t \approx 600$). For $t \gtrsim 520$ the rolls originating upstream of the B\"odewadt layer adopt a larger wavelength, a chaotic dynamics with no exact recurrence, and their radial phase speed is reduced (in absolute value) compared to the edge dynamics. Past $r \approx 4.5$ they decelerate towards the axis with a shorter wavelength and disappear for $r \le 1.5$. The most energetic part of the stationary chaotic regime is found for $4<r<6$ consistently with observations from Fig. \ref{evc}c. This is the region where the bulk azimuthal profile departs most dramatically from {the self-similar profile}, and where ejections from the stator boundary layer occur. }

\subsection{{Saddle-node bifurcation} and mesh dependence}

Figure~\ref{fig:edge1} reports the time average of $a(t)$ for the solutions lying on the edge, together with the levels corresponding to the chaotic regime. The top (chaotic) branch and the edge branch approach {each other} for decreasing $Re$ and visually suggest that they might connect for $Re\approx1800$. By analogy with most subcritical shear flows \cite{schneider2009edge}, we speculate that the two branches merge in what could abusively be labelled as a saddle-node bifurcation at $Re=Re_{SN} \approx1800$ (the fact that the top branch is chaotic makes the concept of saddle-node bifurcation somewhat undefined). Interestingly this value of $Re_{SN}$ matches approximately the value reported in Table~\ref{resn} at which the periodic subcritical branch folds back. This further highlights the connection between that branch and the edge branch. 

The numerical edge tracking was conducted using the two different grid resolutions R0 and R2. By focusing again on the observable $a$, better spatial convergence is observed for the edge branch than for the top branch in Figure~\ref{fig:edge1}. This is a common observation in the shear flows: the edge state generally needs lower resolution than the corresponding turbulent solutions \cite{wang2007lower}. {Despite slight differences both branches can be claimed to be qualitatively well captured numerically.} 
\begin{figure}
    \centering
        \vspace{2mm}
    \includegraphics[width=0.5\columnwidth]{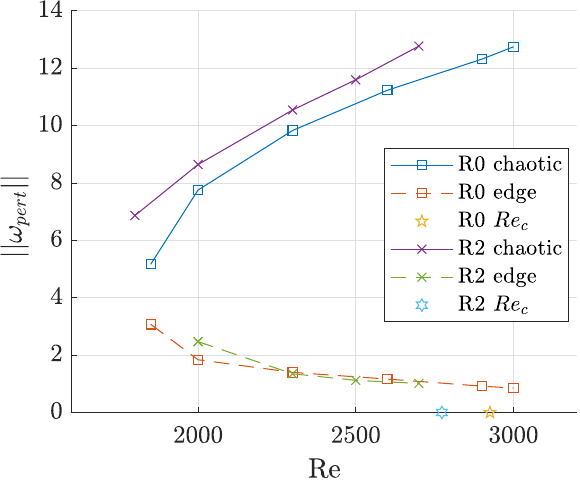}
    \caption{Bifurcation diagram $a(Re)$. Edge state branch and chaotic branch obtained using time integration. Spatial resolutions R0 and R2. The critical value of $Re_c$ corresponding to each  resolution is shown using stars.}
    \label{fig:edge1}
\end{figure}

\section{Summary and outlook} \label{sec-summary}

In this work the axisymmetric rotor-stator flow {has been} revisited, with the aim to explain the dynamical origin of the chaotic rolls reported in experiments and numerics for $Re$ as low as 180 \cite{Gauthier_jfm_1999}. {Former works by other authors suggested that the rolls could be a direct response of the {rotor-stator system} to {imperfections inherent to} the experiments, such as noise of permanent disturbances. Although we do not dismiss the possible role of external perturbations, we adopted here the point of view of bifurcation theory and looked instead for nonlinear branches of \emph{self-sustained} solutions.} The various branches of nonlinear solutions found contribute to a better global picture of the axisymmetric dynamics.\\

It was {first} shown using standard linear stability analysis that the axisymmetric base flow loses stability at a finite Reynolds number $Re_c\approx2900$. At $Re_c$ a Hopf branch of supercritical solutions emerges. {This branch, difficult to continue numerically, was successfully captured using {a Harmonic Balance Method initialised by SCM}, a method free from the use of timesteppers. The associated branch and its folds appear in all cases restricted to a very narrow $O(1)$ interval of values of $Re$, and does not explain the subcriticality of the chaotic rolls.} \\

Even though the bifurcation from the base flow appears supercritical, subcritical branches {of axisymmetric finite-amplitude solutions have been identified}. One of them is a branch of exactly periodic solutions, the other one is a branch of edge states separating the laminar and turbulent regimes. {These two branches appear connected through a bifurcation scenario to be determined.} {They are unstable and exist down to $Re=Re_{SN} \approx 1800$. On one hand, this value of $Re_{SN}$ is about $40\%$ lower than the critical Reynolds number $Re_c$, which justifies the existence of subcritical rolls at least within this $Re$-interval. On the other hand, this value of $Re_{SN}$ is not low enough to explain the experimental observations.} 

\begin{figure}    
\centering
        \vspace{2mm}
    \includegraphics[width=0.5\columnwidth]{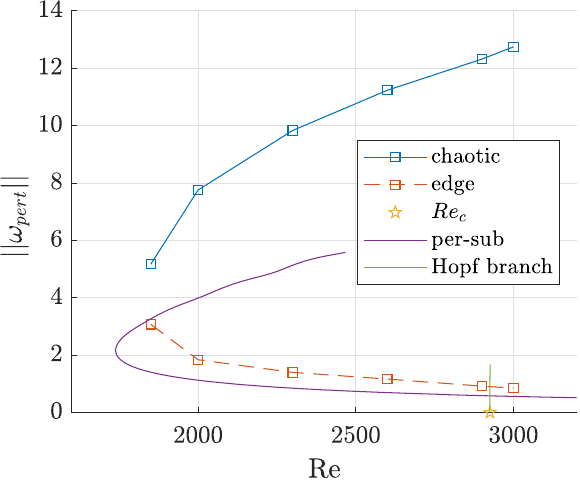}
    \caption{Summary bifurcation diagram $a(Re)$. Three different branches are shown : edge state branch together with the chaotic branch (red/blue squares), periodic-subcritical branch (purple solid line), and the Hopf branch arising at $Re_c$ (green solid line). Spatial resolution R0. }
    \label{summary-graph}
\end{figure}


{The current results {clarify} the global \emph{nonlinear} picture for the transition to chaos in axisymmetric rotor-stator flow. Branches of nonlinear invariant solutions exist in the range $Re=1800-3000$ and were identified numerically using various techniques. Whether stable or unstable, they have little connection to experimentally observed rolls observed experimentally for $Re=200$.
{The exact numerical values for the thresholds of each interval are {indicative but were found} to depend {moderately} on the numerical resolution. The general picture is independent of the mesh resolution.} The $Re$-range can hence be conceptually divided into four separate regions as in Figure \ref{fig:last} :}
\begin{itemize}
    \item I - ($0 \le Re \lesssim 200$) {The base flow is stable and no asymmetric sustained instability mode can occur} neither in the experiments nor in the numerics. 
    \item II - ($Re=200-1800$) The flow is stable both linearly and nonlinearly to axisymmetric perturbations. Sustained axisymmetric rolls can be observed in the experiments, but only as the result of continuous forcing by external perturbations. If such a forcing disappears, the flow relaminarises. \review{Some unstable or stable solutions with small attraction basin may exist but have not been found and, if any, are believed to play limited role in the dynamics.} 
    \item III - $(1800 \lesssim Re \lesssim Re_c \approx 3000)$ High-amplitude branches of self-sustained solutions exist and {contribute to the formation of a chaotic set, possibly a chaotic attractor. Although the base flow} is still linearly stable, strong enough perturbation can trigger chaotic rolls in the flow, linked to the existence of high-amplitude solutions. {For infinitesimally small disturbance levels the system still responds to external excitations as if there were no finite-amplitude solutions.} 
    \item IV - ($Re>Re_c \approx 3000$) The base flow is linearly unstable to axisymmetric perturbations. {Infinitesimal} perturbations to that base flow lead to unsteady rolls.
\end{itemize}

\begin{figure}    
\centering
        \vspace{2mm}
    \includegraphics[width=\columnwidth]{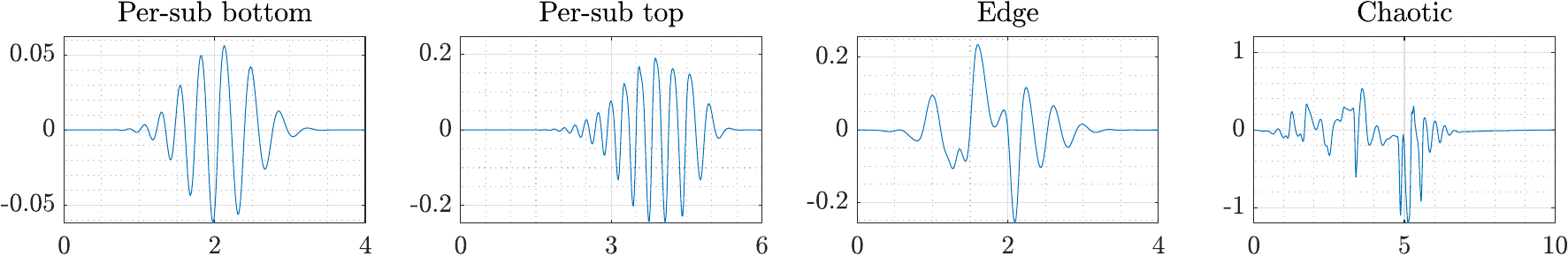}
    \caption{Instantaneous radial profiles of $u_{\theta}(r)$ at $z=0.9375$ for various solutions found at $Re=2300$. From left to right : Periodic subcritical bottom branch, periodic subcritical top branch, Edge state and chaotic branch. Spatial   resolution R0. The horizontal and vertical scales vary from subfigure to subfigure. 
    }
    \label{summary-graph2}
\end{figure}

\begin{figure}
    \centering
    \includegraphics{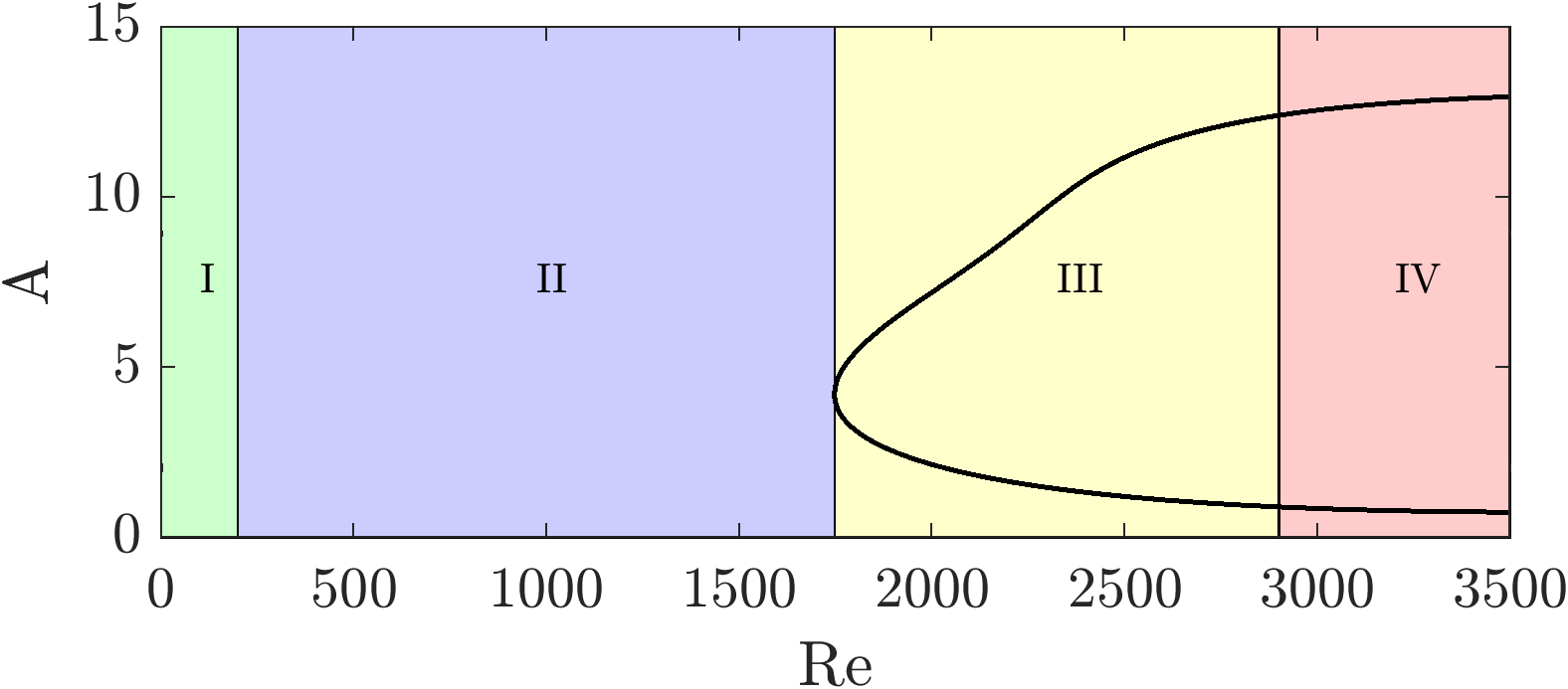}
    \caption{{Conclusion sketch from this study. Four different dynamical regions for axisymmetric rolls have been identified : I)~unconditionally stable base flow, II)~the base flow is stable but rolls appear as response to external forcing, III)~the base flow is only linearly stable but unstable to some finite-amplitude perturbations, chaotic rolls can be sustained without forcing and IV)~chaotic rolls are sustained by the linear instability of the base flow. The solid black lines denotes finite-amplitude solutions described in this paper.}}
    \label{fig:last}
\end{figure}

The branches of axisymmetric solutions found in the present work are summarised in {the bifurcation diagram of} Figure \ref{summary-graph}. In Figure \ref{summary-graph2} {the} individual solutions are {depicted} using a representative instantaneous radial profile of azimuthal perturbation $u_{\theta}(r,z=0.9375)$. 
{All profiles have in common the presence of a spatially growing wave in the decreasing radial direction, as a remains from the convective instability reported at smaller $Re$.} {The profiles from 
the periodic subcritical solution correspond to a simple localised wavepacket localised in the self-similar region. They are all qualitatively similar to the critical eigenvector from Fig. \ref{evc}, but the exact location of their geometric center within the B\"odewadt layer varies. The radial profile of the edge state is spatially less organised than the previous wavepackets, it remains spatially localised but is now characterised by larger wavelengths. By comparison, the chaotic rolls (rightmost frame) have similar amplitude levels in the region closest to the axis but a more energetic core around $r \approx 5$, consistently with the description in Subsection \ref{instab:edge}.}

{The solutions from the Hopf branch, initially believed to be important for the subcritical dynamics, are restricted to a very narrow interval. They do not appear clearly connected to the other solutions and do not contribute to the origin of the chaotic rolls. The dynamical connections between the other different states found in this work are conjectured as relatively simple. The periodic subcritical states appear, upon increasing $Re$, in a saddle-node bifurcation. They do not bifurcate from the base flow, {at least not in the range of $Re$ investigated}. The edge state branch results from one (or several) bifurcation(s) of the lower-branch of periodic subcritical states. The edge state is finally connected directly to the chaotic rolls via its linear instability. The fact that the whole subcritical dynamics appears fully disconnected dynamically from the base flow is common in open shear flows, it was also reported in rotor-stator flow by Lopez {\it et al.} for $\Gamma=5$ \cite{Lopez_pof_2009}, for which no axisymmetric instability of the base flow was identified. The presence for $\Gamma=10$ of a critical point with a linear stability apparently leaves this picture unchanged. A radical difference between $\Gamma=5$ and $10$ is the more complex dynamics observed for higher $\Gamma$. This is consistent with the qualitative sketch that wave-like perturbations, once born in the corner/shroud region, are advected radially inwards along the stator, and that the further they can travel from their origin, the more they get amplified and the more chaotic they become. 
}

Natural extensions of the current work would concern in priority the intervals in $Re$ labelled II and III.} A possibility would to be to conduct, in the spirit of \cite{do2010optimal}, numerical simulations forced by noise, yet at lower values of $Re$.
The exact role of the non-normality of the evolution operator linearised around the base flow deserves a deeper understanding, and could be quantified using resolvent analysis \cite{Trefethen_science_1993}. 
{Besides, the dynamical role of the inertial waves present in the core of the cavity, their contribution to the transition process and to the individual nonlinear solutions deserve more investigation.}
Although the present study was focused on the influence of $Re$ for a given geometry, the influence of the aspect ratio $\Gamma=R/H$ of the rotor-stator cavity {also needs} be addressed. In particular, the influence of $\Gamma$ on the linear stability threshold $Re_c$ deserves more investigation. 
Eventually, experiments and numerics have shown that three-dimensionality can generally not be avoided due to the linear instability of the axisymmetric base flow to spiral modes. In the same spirit as the current study, the identification of unstable three-dimensional nonlinear solutions would be a relevant way, by reconstructing the state space of the system \cite{gibson2008visualizing}, to investigate the turbulent states observed in practice. These suggestions for future investigation are left for forthcoming studies.


\section*{Acknowledgments}

The authors would like to thank L.S. Tuckerman for constructive discussions and E. Sleimi for the preliminary work on the subject as a part of his internship.

\appendix*


\section{Threshold sensitivity to the numerical resolution}

The sensitivity of the critical Reynolds number is addressed {with respect to the numerical mesh and to the way the singular boundary conditions are treated numerically}. Since the rotating shroud wall meets the stationary stator disc a discontinuity of the boundary condition for the azimuthal velocity is present in the associated corner. Such discontinuities are known to deteriorate the accuracy of spectral simulations, which has promoted the use of regularisation techniques used in \cite{Lopez_pof_2009,Serre_pof_04}.

\review{The sensitivity of the instability threshold is tested by using a series of meshes summarised in Table \ref{reolutions} and illustrated in Figure \ref{fig:recmesh}, which reports the value of $Re_c$ for each spatial discretisation.}
Additionally, the influence of {\emph{regularizing}} the corner singularity is also considered. This is achieved {here} by smoothing out the boundary condition at the B\"odewadt corner, imposing an exponential velocity profile of the form 
\review{$u_{\theta}=r\exp{\left(\frac{r-\Gamma}{\varepsilon}\right)}$.} 
\review{Two regularisations have been considered : $\varepsilon=0.003$ and $\varepsilon=0.006$. The case without any regularization ($u_{\theta}=0$) is referred to as $\varepsilon=0$ for ease of notation.}

The spatial discretisation of the governing equations and boundary conditions is {second order accurate}. It can be therefore expected that {the value of $Re_c$ also converges with the {same order as} the mesh is refined. The order of convergence for $Re_c$ is estimated based on three different resolutions R2, R3 and R4 as}
\begin{equation}
\textrm{order}=\log_{1.5}\left(\frac{Re_{c,R3}-Re_{c,R2}}{Re_{c,R4}-Re_{c,R3}}\right)
\end{equation}
The choice of the base 1.5 for the logarithm reflects the ratio between {consecutive} grids in Table \ref{reolutions}. As also noted in the Table, the order of convergence is close to {2.0}, irrespective of the regularisation used. Smoothing out the discontinuous boundary condition, although it affects quantitatively $Re_c$, does not affect the order of convergence. {The dependence of $Re_c$ on the regularisation is due to the modification of the base flow} {linked to the modified boundary condition} on the stator and not {to the removal of the} singularity. {We conclude} that using a discontinuous boundary condition does not deteriorate the results of the finite difference approximations used throughout this study. Similar conclusions were drawn for the lid-driven cavity, which also displays corner singularities, see e.g. the recent review~\cite{kuhlmann_2019}. {As a compromise} the spatial resolution R2 with no corner regularisation will provide meaningful results while requiring a manageable computational effort. As a computationally cheaper compromise spatial resolution R0 will also be extensively used. {Both corresponding meshes are highlighted} in bold in Table~\ref{reolutions}.

\begin{table}[h]
\centering
\begin{tabular}{l|c|c|c|c|lll}
resolution & $N_r$ & $N_z$ &type&DOF&$\varepsilon=0$&$\varepsilon=0.003$&$\varepsilon=0.006$ \\ \hline
{\textbf{R0}}  & \textbf{600}&\textbf{160}&uniform& \textbf{390 k} &\textbf{2925.47}& & \\
R1 & 683&128&non-uniform& 356 k & 2963.41 & 2985.43 & 3035.61 \\
 {\textbf{R2}} & {\textbf{1024}}&\textbf{192} &non-uniform&\textbf{796 k} & {\textbf{2773.3}}  & 2789.01 & 2826.55 \\
R3 & 1536&288&non-uniform&1.7 m & 2697.48 & 2711.71 & 2746.33 \\
R4 & 2304&432&non-uniform&4.0 m  & 2663.96 & 2677.9  & 2711.97 \\
R5 & 3456&648&non-uniform&9.0 m & 2648.9  & 2662.8  & 2696.8 \\ \hline
&&&& extrapolation & 2636.61   &    2650.59      &    2684.81 \\ \hline
&&&& order &2.01	&2.04	&2.09 \\
\end{tabular}
\caption{Critical Reynolds number $Re_c$ depending on the the spatial discretisation. From R1 to R5 the ratio between two consecutive grid resolutions is 1.5 in each direction.}
\label{reolutions}
\end{table}
\begin{figure}[h]
    \centering
    \includegraphics[width=0.49\textwidth]{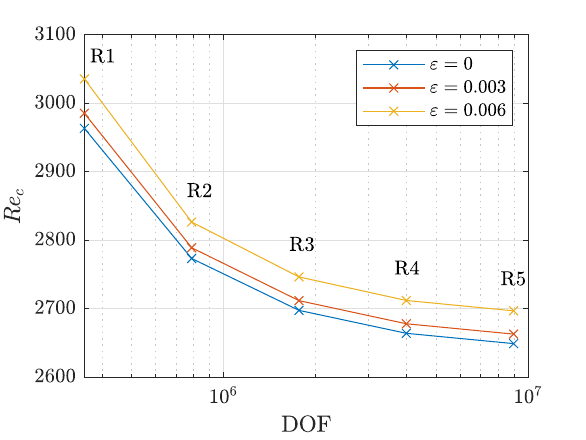}
    \includegraphics[width=0.49\textwidth]{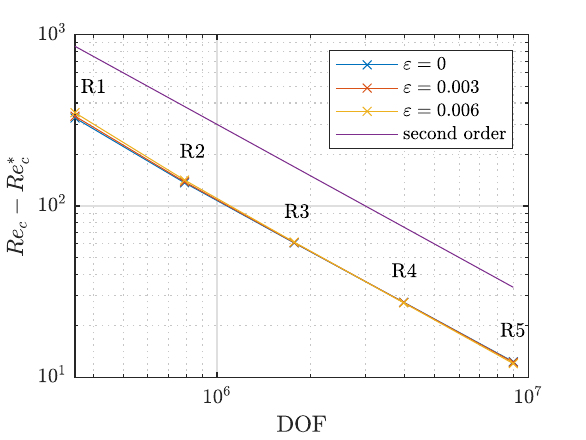}
    \caption{\review{Left: critical Reynolds number $Re_c$ depending on the the spatial discretisation in the case of regularised and non-regularised corner singularity. Right: absolute error calculated using the Richardson extrapolated value $Re_c^*$ compared with a second order convergence curve.}}
    \label{fig:recmesh}
\end{figure}

\newpage

\end{document}